\documentclass[prd]{revtex4}

\usepackage{amsmath,bm}
\usepackage{graphicx}

\begin{document}

\title{Non-contact gears: I.
Next-to-leading order contribution to lateral Casimir force between 
corrugated parallel plates.}
\author{In\'{e}s Cavero-Pel\'{a}ez}
\email{cavero@spectro.jussieu.fr}
\affiliation{Laboratoire Kastler Brossel,
Universit\'{e} Pierre et Marie Curie, ENS, CNRS, Campus Jussieu, 
University Paris 6, Case 74, F-75252 Paris, cedex 05, France.}
\author{Kimball A. Milton}
\email{milton@nhn.ou.edu}
\homepage{http://www.nhn.ou.edu/
\author{Prachi Parashar}
\email{prachi@nhn.ou.edu}
\author{K. V. Shajesh}
\email{shajesh@nhn.ou.edu}
\homepage{http://www.nhn.ou.edu/
\affiliation{Oklahoma Center for High Energy Physics
and Homer L. Dodge Department of Physics and Astronomy,
University of Oklahoma, Norman, OK 73019, USA.}

\date{\today}

\begin{abstract}
We calculate the lateral Casimir force between corrugated 
parallel plates, described by $\delta$-function potentials,
interacting through a scalar field, using the multiple scattering 
formalism. The contributions to the Casimir energy due to 
uncorrugated parallel plates is treated as a background from
the outset. We derive the leading- and next-to-leading-order 
contribution to the lateral Casimir force for the case when the 
corrugation amplitudes are small in comparison to corrugation
wavelengths. We present explicit results in terms of 
finite integrals for the case of the Dirichlet limit, and exact 
results for the weak-coupling limit, for the leading- and 
next-to-leading-orders. The correction due to the 
next-to-leading contribution is significant. In the weak 
coupling limit we calculate the lateral Casimir force exactly 
in terms of a single integral which we evaluate numerically.
Exact results for the case of the weak limit allows us to estimate 
the error in the perturbative results. We show that the error 
in the lateral Casimir force, in the weak coupling limit, when 
the next-to-leading order contribution is included is remarkably
low when the corrugation amplitudes are small in comparison to 
corrugation wavelengths. We expect similar conclusions to hold for 
the Dirichlet case. The analogous calculation for the 
electromagnetic case should reduce the theoretical error 
sufficiently for comparison with the experiments.
\end{abstract}

\maketitle

\section{Introduction}

The Casimir force, as exhibited between neutral metallic parallel plates,
was discovered theoretically in 1948~\cite{Casimir:1948dh}.
The Casimir torque between asymmetric materials was first studied in
1973~\cite{Barash:1973}.
Recently, theoretical study of the lateral Casimir force between corrugated 
parallel plates was pioneered and developed by the MIT group in
\cite{Golestanian:1997ks, Golestanian:1998,
      Emig:2001dx, Emig:2003, Buscher:2004tb}.
In particular, in \cite{Emig:2003}, the authors evaluated analytic 
expressions for the lateral Casimir force, to the leading order,
between two corrugated parallel plates perturbatively.
Experimentally, the Casimir interaction between 
corrugated surfaces was explored during the same period of time
by Roy and Mohideen in~\cite{Roy:1999}. This experiment measured 
the lateral Casimir force between a plate, with small sinusoidal 
corrugations, and a large sphere with identical corrugations. 
The motivation was to study the nontrivial boundary 
dependence in the Casimir force. 
The experimental parameters in our notation are (see figure~\ref{corru}): 
$h = 60~\text{nm}$,
$d = 1.1~\mu\text{m}$, and
$a = 0.1-0.9~\mu\text{m}$,
where $h$ is the height of the corrugations, $d$ is the wavelength 
of the corrugations, and $a$ is the mean distance between the plates.
The corresponding dimensionless quantities are:
$k_0a = 0.6 - 5.1$,
$\frac{h}{a} = 0.07 - 0.6$, and
$k_0h = 0.3$,
where $k_0$ is the wavenumber related to the spatial wavelength 
of the corrugations.

Experimental data was analyzed based on the 
theoretical results obtained from the proximity force approximation (PFA), 
and has been presented in~\cite{Chen:2002,Chen:2002b}.
The validity of the PFA in the above analysis 
has been the topic of a recent debate and controversy,
see \cite{Rodrigues:2006, Chen:2007, Rodrigues:2007}.
Theoretical results based on perturbative approximations as
done in \cite{Emig:2001dx} do not settle the issue because the
error keeping only the leading order may be high. It is generally believed
that the next-to-leading-order calculation will be able to throw
light on the issue. We carry out this calculation for the 
case of scalar fields. The analogous calculation for the 
electromagnetic case should now be straightforward.

This paper in principle is an offshoot of~\cite{Cavero:}
where we shall deal with corrugated cylinders to study non-contact gears.
While evaluating the leading order for the case of corrugated cylinders 
it was noticed that it would be possible to extend the calculation to the 
next-to-leading order. This led to the study in the present paper. 
In this installment we present the next-to-leading-order
calculation for the case of corrugated parallel plates. 
The leading order calculation for the corrugated cylinders,
which in itself is a significant result, will form the 
sequel~\cite{Cavero:} of this paper.
The next-to-leading-order calculation for the corrugated cylinders
is in progress. 

\section{Formalism}

In this section we shall describe the formalism and derive the 
key formula used for calculating the Casimir energy.
This has been done in various papers before,
(see~\cite{Emig:2007me},~\cite{Kenneth:2006vr},~\cite{Milton:2007wz}, 
and references therein).
We hope our derivation using Schwinger's quantum action principle
techniques will be illuminating.
In an earlier paper~\cite{Milton:2007wz} describing the 
multiple scattering formalism it was mentioned that the 
use of the scattering matrix, $T$, was equivalent to using the 
full Green's function, and required the same computational effort.
As a justification of this comment we exclusively use the full 
Green's function in this article.

\subsection{Vacuum energy in field theory}

Let us consider a scalar field, $\phi(x)$, interacting with 
a scalar background potential, $V(x)$, described by the Lagrangian
density
\begin{equation}
{\cal L}(\phi (x))
= -\,\frac{1}{2} \,\partial_\mu \phi (x) \partial^\mu \phi (x)
  - \frac{1}{2} \,V(x) \, \phi (x)^2.
\end{equation}
In terms of the source function, $K(x)$, corresponding to the 
scalar field, we write the action for this description to be
\begin{equation}
W[\phi;K] = \int d^4x \Big[ K(x)\phi(x) + {\cal L}(\phi(x)) \Big].
\label{wpk}
\end{equation}
The vacuum to vacuum persistence amplitude,
\begin{equation}
Z[K] = \langle 0_+|0_- \rangle^K,
\end{equation}
which generates the amplitudes for all the physical processes,
satisfies Schwinger's quantum action principle,
\begin{equation}
\delta Z[K] = i\,\langle 0_+|\,\delta W[\phi;K] \,|0_- \rangle^K.
\label{QAP}
\end{equation}
Our immediate task will be to get a formal solution for the 
vacuum amplitude, $Z[K]$, in the form
\begin{equation}
Z[K] = e^{i \,W[K]},
\label{z=eiw}
\end{equation}
where $W[K]$, which is not an action anymore, is dependent only on the 
source function. Note that the action, $W[\phi;K]$ in eq.~\eqref{wpk},
which satisfied the action principle, was described in terms 
of both the (operator) scalar field and the source function.

Variation with respect to the source function in the quantum 
action principle in eq.~\eqref{QAP} allows us to write 
\begin{equation}
\varphi(x) \equiv \frac{\langle 0_+|\,\phi(x)\,|0_- \rangle^K}{Z[K]}
= \frac{1}{Z[K]} \frac{1}{i} \frac{\delta Z[K]}{\delta K(x)},
\label{eff-field}
\end{equation}
where the redefined scalar field, on the left of the above expression,
is an effective field.
This can be used to replace operator fields with functional derivatives 
with respect to the sources.
Variation with respect to the scalar field in eq.~\eqref{QAP} gives us
\begin{equation}
- \Big[ \partial^2 - V(x) \Big]
\, \frac{1}{i} \frac{\delta Z[K]}{\delta K(x)} = K(x) Z[K],
\label{ginz=kz}
\end{equation} 
which can be immediately inverted and written in the form, 
after using eq.~\eqref{eff-field},
\begin{equation}
\varphi (x) =
\frac{1}{Z[K]} \frac{1}{i} \frac{\delta Z[K]}{\delta K(x)}
= \int d^4 x^\prime G(x,x^\prime) K(x^\prime), 
\label{zinz=gz}
\end{equation}
where we defined the inverse of the differential operator,
the Green's function, as 
\begin{equation}
- \Big[ \partial^2 - V(x) \Big] \, G(x,x^\prime) = \delta^{(4)} (x-x^\prime).
\label{green}
\end{equation}
The solution to eq.~\eqref{zinz=gz} is a Gaussian in the functional 
sense, and when written in the desired form in eq.~\eqref{z=eiw},
lets us identify 
\begin{equation}
W[K] = Q[V] + \frac{1}{2} \int d^4 x \int d^4 x^\prime
K(x) G(x,x^\prime) K(x^\prime),
\end{equation}
where $Q[V]$ is a functional of the background potential alone.
For the case when the background potential is switched off, described 
by $V=0$, we have $Z_0[K] = \text{exp}(i W_0[K])$, where
\begin{equation}
W_0[K] = Q[0] + \frac{1}{2} \int d^4 x \int d^4 x^\prime
K(x) G_0(x,x^\prime) K(x^\prime),
\label{free-act}
\end{equation}
where, $G_0(x,x^\prime)$ is the corresponding Green's function 
in eq.~\eqref{green} for the case when $V=0$. Now, in the absence of
a source function the vacuum should not decay, 
which amounts to the statement $\langle 0_+|0_- \rangle^{K=0} = 1$.
This implies that $W_0[0] = 0$ which when used in eq.~\eqref{free-act}
lets us conclude that $Q[0] = 0$.

Variation with respect to the background potential in eq.~\eqref{QAP} yields
\begin{equation}
\delta Z[K] = - \frac{i}{2} \int d^4x \, \delta V(x)
\,\langle 0_+| \,\phi(x) \phi(x)\, |0_- \rangle,
\end{equation}
where we can replace the operator fields with functional derivatives
and then write the formal solution as 
\begin{equation}
Z[K] = e^{- \frac{i}{2} \int d^4x \, V(x)
  \frac{1}{i} \frac{\delta}{\delta K(x)} 
  \frac{1}{i} \frac{\delta}{\delta K(x)} }
e^{\frac{i}{2} \int d^4 x \int d^4 x^\prime
K(x) G_0(x,x^\prime) K(x^\prime)}.
\end{equation}
Using the standard identity, we have
\begin{equation}
W[K] = - \frac{i}{2} \,\text{Tr} \,\text{ln} \,G G_0^{-1} 
+ \frac{1}{2} \int d^4 x \int d^4 x^\prime
K(x) G(x,x^\prime) K(x^\prime),
\label{action}
\end{equation}
which uses $(1+G_0V)^{-1}G_0 = G$. 
Using eq.~\eqref{action} in eq.~\eqref{z=eiw}
we observe that in the presence of a background the vacuum to
vacuum transition amplitude is not unity, instead it evaluates to 
\begin{equation}
\langle 0_+|0_- \rangle^{K=0} = e^{i W[0]}
= e^{\frac{1}{2} \,\text{Tr} \,\text{ln} \,G G_0^{-1}}.
\label{z=iw0}
\end{equation}

For the case when the process under investigation is time independent,
denoting $\tau$ to be the time associated with the physical process,
we can evolve the vacuum state using the Hamiltonian of the system
and thus conclude
\begin{equation}
\langle 0_+|0_- \rangle^{K=0}
= \langle 0_-| e^{-i H \tau} |0_- \rangle
= e^{- i E \tau},
\label{z=iet}
\end{equation}
where we assumed the vacuum to be an eigenstate of the Hamiltonian.
Comparing the two forms, eq.~\eqref{z=iw0} and eq.~\eqref{z=iet},
we thus identify the energy of the vacuum in the presence of a 
background to be 
\begin{equation}
E = \frac{i}{2 \tau} \,\text{Tr} \,\text{ln} \,G G_0^{-1}
\label{E=TrLnGG}
\end{equation}
which serves as the central formula for calculating the Casimir energy.
Further, for the case of a time independent situation,
making a Fourier transformation in the time variable and 
using translational independence in time, we can formally write
\begin{equation}
E = - \frac{i}{2} \int \frac{d\omega}{2\pi} 
\,\text{Tr} \,\text{ln} 
\Big[ - \omega^2 - \nabla^2 + V \Big] 
+ \frac{i}{2} \int \frac{d\omega}{2\pi}
\,\text{Tr} \,\text{ln} 
\Big[ - \omega^2 - \nabla^2 \Big] 
\end{equation}
where we used eq.~\eqref{green} to define the Fourier transformed
Green's function in the form 
$\bar{G}^{-1} = [ - \omega^2 - \nabla^2 + V ]$.
Integrating by parts and throwing away the boundary terms we derive 
\begin{equation}
E = - \frac{i}{2} \int \frac{d\omega}{2\pi} 2 \omega^2 \,\text{Tr} \,\bar{G} 
+ \frac{i}{2} \int \frac{d\omega}{2\pi} 2 \omega^2 \,\text{Tr} \,\bar{G}_0, 
\end{equation}
which is an alternate expression for evaluating the Casimir energy.
This latter form has been used in studies~\cite{Milton:2004vy}
related to surface divergences 
which requires the evaluation of the energy density rather than 
the total energy.

\subsection{Statement of the problem}

We consider two semitransparent, 
corrugated plates, parallel to the $x-y$ plane,
described by the potentials,
\begin{equation}
V_i(z,y) = \lambda_i \,\delta (z - a_i - h_i(y)),
\label{pot-i}
\end{equation}
where $i=1,2$, are labels that identify the individual plates,
and we define $a = a_2 - a_1 > 0$.
The functions $h_i(y)$ are designed to describe the 
corrugations associated with the individual plates.
Let us define the function
\begin{equation}
a(y) = a + h_2(y) - h_1(y)
\label{a-th}
\end{equation}
which measures the relative corrugations between the two plates.
We shall define the corrugations $h_i(y)$ so that 
the mean of the relative corrugations is $a$.
In general we require $a(y)>0$, which is a restriction on the corrugations.
Translational invariance is assumed in the $x$ direction.

The change in energy due to the change in the mean distance between 
the plates, $a$, leads to the conventional Casimir force 
which points in the direction perpendicular to the surface of the plates,
and is expressed as
\begin{equation}
F_\text{Cas} = - \frac{\partial E}{\partial a},
\label{casfE}
\end{equation}
where $E$ is the total Casimir energy associated with the
corrugated plates including the divergent contributions
associated with the single plates.
The divergent contributions being independent of $a$,
do not contribute to the Casimir force.
When the plates have corrugations on them we expect to have
a change in the total energy due to a shift of one of the plates
parallel (lateral) to the other plate.
The force corresponding to the change in the total energy due to this 
shift acts in the lateral direction and is called the lateral 
Casimir force. The shift is mathematically described by a translation
in the $y$-coordinate, $h_1(y + y_0)$, which corresponds to a 
lateral shift of one plate with respect to the other.
The lateral force is expressed as
\begin{equation}
F_\text{Lat} = - \frac{\partial E}{\partial y_0},
\label{latfE}
\end{equation}
where $y_0$ is the measure of the translational shift.

We note that there will be no lateral force between the 
plates if the corrugations are switched off by setting
$h_i(y) = 0$, $i=1$ and $2$. The physical quantities associated
with this configuration will thus act as a background, and a reference,
and we shall find it convenient to denote them by the superscript $(0)$
to mean zeroth order. The potential for the background 
will thus be described by
\begin{equation}
V_i^{(0)}(z) = \lambda_i \,\delta (z - a_i),
\end{equation}
which has no $y$ dependence.
The total Casimir energy associated with the background,
due to the two uncorrugated plates, will be denoted as $E^{(0)}$,
which will include the divergent contributions associated with the 
single plates. 
This background energy will be independent of the displacement
$y_0$ due to the $y$ independence
of this configuration. Thus, we can conclude that 
\begin{equation}
F^{(0)}_\text{Lat} = - \frac{\partial E^{(0)}}{\partial y_0} =0.
\label{latfE0}
\end{equation}
We further note that there will be no lateral force between the plates 
if either one of the plates have their corrugations switched off
by setting $h_i(y) = 0$, $i=1$ or $2$. 
The Casimir energy associated with this configuration can be written as
\begin{equation}
E_i = E - E^{(0)}
\end{equation}
where $E_i$ is the additional contribution to the energy with respect
to the background energy due to the presence of the corrugations
on one of the plates.
Throughout this article we shall use $\Delta$ to mean the deviation
from the background. For example, we will have
$\Delta V_i(z,y) = V_i(z,y) - V_i^{(0)}(z)$.
Using the argument of $y$ independence, or translational symmetry,
we can again conclude that
\begin{equation}
(F_\text{Lat})_i = - \frac{\partial E_i}{\partial y_0} =0.
\end{equation}

In light of the above observations we are led to write the 
total Casimir energy, for the case when both the plates have 
their corrugations switched on, in the form
\begin{equation}
\Delta E = E - E^{(0)}(a) = E_1(a,h_1) + E_2(a,h_2) + E_{12}(a,h_1,h_2,y_0),
\label{DE=DEi+}
\end{equation}
where $E_{12}$ is the contribution to the total energy due to the 
interaction between the corrugations in the plates. Only this 
part of the total energy contributes to the lateral Casimir force. 
Thus, we conclude that
\begin{equation}
F_\text{Lat} = - \frac{\partial}{\partial y_0} \Delta E
= - \frac{\partial E_{12}}{\partial y_0}.
\label{latf}
\end{equation}
Our central problem will be to evaluate $E_{12}$ 
using eq.~\eqref{E=TrLnGG}.

\subsection{Casimir energy contributing to the lateral force}

Using the central formula derived in the multiple scattering formalism
in eq.~\eqref{E=TrLnGG} to evaluate eq.~\eqref{DE=DEi+} we have
\begin{equation}
\Delta E = E - E^{(0)} 
= \frac{i}{2 \tau} \,\text{Tr} \,\text{ln} \,G G_0^{-1}
  - \frac{i}{2 \tau} \,\text{Tr} \,\text{ln} \,G^{(0)} G_0^{-1}
= \frac{i}{2 \tau} \,\text{Tr} \,\text{ln} \,G {G^{(0)}}^{-1},
\label{DE=trln}
\end{equation}
where $G_0$ is the free Green's function introduced in eq.~\eqref{free-act}.
Note that $G_0$ cancels in the above expression 
and the reference is now with respect to the uncorrugated plates. 
The Green's function $G$ satisfies eq.~\eqref{green} 
with potentials described by eq.~\eqref{pot-i},
which in symbolic notation reads 
\begin{equation}
\Big[ -\partial^2 + V_1 + V_2 \Big] G = 1,
\end{equation}
and the corresponding Green's function associated with the 
background satisfies the differential equation,
\begin{equation}
\Big[ -\partial^2 + V_1^{(0)} + V_2^{(0)} \Big] G^{(0)} = 1.
\label{G0eqn}
\end{equation}
The above two equations can be used to deduce
\begin{equation}
G {G^{(0)}}^{-1} = \left[ 1 + G^{(0)}\Delta V_1 + G^{(0)}\Delta V_2\right]^{-1}.
\end{equation}
Next, one makes the observation that 
the above expression can be rewritten in the form
\begin{equation}
G {G^{(0)}}^{-1}
= G_2 {G^{(0)}}^{-1} \Big[ 1 - G_1 \Delta V_1 G_2 \Delta V_2 \Big]^{-1}
  G_1 {G^{(0)}}^{-1},
\label{GG0-1}
\end{equation}
where $G_i$ ($i=1,2$) are the Green's functions for the 
parallel plates when only one of the plates has corrugations on it.
The differential equations for $G_i$'s are
\begin{equation}
\left[-\partial^2 + V_1^{(0)} + V_2^{(0)} + \Delta V_i\right]G_i = 1,
\end{equation}
which together with eq.~\eqref{G0eqn} can be used to deduce
\begin{equation}
G_i {G^{(0)}}^{-1} = (1 + G^{(0)}\Delta V_i)^{-1}
= 1 - G^{(0)}\Delta V_i (1 + G^{(0)}\Delta V_i)^{-1}
= 1 - (1 + G^{(0)}\Delta V_i)^{-1} G^{(0)}\Delta V_i.
\label{GiG0}
\end{equation}
Using eq.~\eqref{GG0-1} in eq.~\eqref{DE=trln} we
we immediately obtain eq.~\eqref{DE=DEi+}. 
The last term in eq.~\eqref{DE=DEi+}, $E_{12}$,
which is the only term that contributes to the lateral force between the 
two corrugated plates, can be read out from eq.~\eqref{GG0-1} 
to be given by the expression
\begin{equation}
E_{12} = - \frac{i}{2\tau} \,\text{Tr} \,\text{ln}
\Big[ 1 - G_1 \Delta V_1 G_2 \Delta V_2 \Big].
\label{DE12}
\end{equation}
We could have written this down immediately, but we hope it was
instructive to go through the steps because it clarifies our notation,
which would anyhow have required us to write many of the above equations,
and also to emphasize that the configuration due to parallel plates 
is treated as a background from the outset.

\subsection{Formal series expansion}

Formally expanding the logarithm in the above expression
and using eq.~\eqref{GiG0} to expand $G_i \Delta V_i$ 
in terms of $G^{(0)} \Delta V_i$
we can write
\begin{equation}
E_{12} = \frac{i}{2\tau} \,\text{Tr} 
\sum_{m=1}^\infty \frac{1}{m}
\Bigg[ 
\sum_{n_1=1}^\infty \sum_{n_2=1}^\infty
(-1)^{n_1} (-1)^{n_2}
\left\{ G^{(0)} \Delta V_1 \right\}^{n_1}
\left\{ G^{(0)} \Delta V_2 \right\}^{n_2}
\Bigg]^m.
\label{formal-series}
\end{equation}
Our potentials in eq.~\eqref{pot-i} can be formally expanded 
in powers of $h_i$ in the form 
\begin{equation}
\Delta V_i(z,y)
= \sum_{n=1}^\infty V_i^{(n)}(z,y)
= \sum_{n=1}^\infty \frac{[-h_i(y)]^n}{n!}
\frac{\partial^n}{\partial z^n} V_i^{(0)}(z)
= \lambda_i \Big[e^{- h_i(y) \frac{\partial}{\partial z}}-1\Big]\delta (z-a_i),
\label{pot-series}
\end{equation}
so we can further write the series expansion as
\begin{equation}
E_{12} = \frac{i}{2\tau} \,\text{Tr}
\sum_{m=1}^\infty \frac{1}{m}
\Bigg[
\sum_{n_1=1}^\infty \sum_{n_2=1}^\infty
(-\lambda_1)^{n_1} (-\lambda_2)^{n_2}
\left\{ G^{(0)} \big[e^{- h_1 \partial} - 1\big]\delta_1 \right\}^{n_1}
\left\{ G^{(0)} \big[e^{- h_2 \partial} - 1\big]\delta_2 \right\}^{n_2}
\Bigg]^m,
\end{equation}
where we again use symbolic notation and suppress the variable dependence
in $h_i$, $\partial$, and the delta functions.

\section{Leading order contribution}
\label{leading-order}

In this section we shall derive the leading order contribution 
to the Casimir energy when the corrugation amplitude is small 
in comparison to the corrugation wavelength. This 
has been evaluated for the Dirichlet scalar case and the 
electromagnetic case in~\cite{Emig:2001dx,Emig:2003}. We 
obtain the result for scalar $\delta$-function potentials as
a warm up exercise in preparation for the calculation for the 
next-to-leading order. In this section we shall illustrate our
methodology which will be further used in the higher
order calculation.

\subsection{Second order perturbation in Casimir energy due to corrugations}

For the particular case when the corrugations can be treated 
as small perturbations we can approximate the potentials
by keeping a few terms in the expansion in eq.~\eqref{formal-series} 
where we use the superscripts $(n)$ to represent the 
$n$-th order perturbation in a quantity.
Thus, to the leading order the interaction energy of the 
corrugations in eq.~\eqref{formal-series} takes the form
\begin{equation}
E_{12}^{(2)} = \frac{i}{2\tau} \,\text{Tr} 
\Big[ G^{(0)} \Delta V_1^{(1)} G^{(0)} \Delta V_2^{(1)} \Big].
\label{dE12-2}
\end{equation}
We observe that the potentials in the $n$-th order are given 
by derivatives acting on $\delta$-functions in eq.~\eqref{pot-series}
which can be transferred over to the Green's functions after 
integration by parts. 
The background Green's function, which is a solution to eq.~\eqref{G0eqn},
can be written as
\begin{equation}
G^{(0)}({\bf x},t,{\bf x}^\prime,t^\prime)
= \int \frac{d\omega}{2\pi} \,e^{-i\omega (t - t^\prime)}
  \int \frac{dk_x}{2\pi} \,e^{i k_x (x - x^\prime)}
  \int \frac{dk_y}{2\pi} \,e^{i k_y (y - y^\prime)}
  \,g^{(0)}(z,z^\prime;\kappa),
\label{G0zzp}
\end{equation}
where $\kappa^2 = \bar{\kappa}^2 + k_y^2$,
with $\bar{\kappa}^2 = k_x^2 - \omega^2$,
where $k_{x,y}$ and $\omega$ are the Fourier variables 
corresponding to the space-time coordinates $x,y$ and $t$ respectively.
The reduced Green's function,
$g^{(0)}(z,z^\prime;\kappa)$, satisfies the equation
\begin{equation}
-\left[ \frac{\partial^2}{\partial z^2} - \kappa^2
  - \lambda_1 \delta (z-a_1) - \lambda_2 \delta (z-a_2) \right]
g^{(0)}(z,z^\prime;\kappa)
= \delta (z-z^\prime).
\label{g0zzp}
\end{equation}

In terms of the reduced Green's function defined above we can write
eq.~\eqref{dE12-2} as
\begin{equation}
\frac{E_{12}^{(2)}}{L_x}
= \int_{-\infty}^{\infty} \frac{dk_1}{2\pi}
  \int_{-\infty}^{\infty} \frac{dk_2}{2\pi}
\,\tilde{h}_1(k_1-k_2) \,\tilde{h}_2(k_2-k_1) 
\,L^{(2)}(k_1,k_2),
\label{DE12=L2}
\end{equation}
where $L_x$ is a large length in the $x$ direction and
where $\tilde{h}_i(k)$ are the Fourier transforms of the functions
$h_i(y)$, which describe the corrugations on the parallel plates,
\begin{equation}
\tilde{h}_i(k) = \int_{-\infty}^{\infty} dy \,e^{-iky} \,h_i(y).
\label{hk=hy}
\end{equation}
The kernel $L^{(2)}(k_1,k_2)$ in eq.~\eqref{DE12=L2} 
is suitably expressed in the form
\begin{equation}
L^{(2)}(k_1,k_2) = - \frac{1}{4\pi} \int_0^\infty \bar{\kappa} \,d\bar{\kappa}
\,\,I^{(2)}(\kappa_1, \kappa_2),
\label{L2=I2}
\end{equation}
where we have switched to imaginary frequencies by a Euclidean 
rotation, $\omega \rightarrow i \zeta$, and defined
\begin{equation}
I^{(2)}(\kappa_1,\kappa_2) = \lambda_1 \lambda_2
\frac{\partial}{\partial z} \,\frac{\partial}{\partial \bar{z}}
       \Big[ \,g^{(0)}(z,\bar{z};\kappa_1)
       \,g^{(0)}(\bar{z},z;\kappa_2)
\Big] \bigg|_{\bar{z}=a_1, z=a_2},
\label{I2}
\end{equation}
where $\kappa_i^2 = \bar{\kappa}^2 + k_i^2$.
Using the reciprocal symmetry in the Green's function 
in eq.~\eqref{I2} we deduce that 
\begin{equation}
I^{(2)}(\kappa_1,\kappa_2) = I^{(2)}(\kappa_2,\kappa_1).
\label{I2sym}
\end{equation}
We evaluate the derivatives in the expression for 
$I^{(2)}(\kappa_1,\kappa_2)$, using the prescription described in
appendix \ref{ddz-green}, as
\begin{eqnarray}
I^{(2)}(\kappa_1,\kappa_2)
&=& - \frac{\lambda_1}{2\kappa_1} \frac{\lambda_2}{2\kappa_2}
\frac{e^{-a(\kappa_1 + \kappa_2)}}{\Delta_1 \Delta_2}
\left[
\kappa_1^2 \Big(1 + \frac{\lambda_1}{2\kappa_1} \Big) 
\Big(1 + \frac{\lambda_2}{2\kappa_1} \Big)
+ \kappa_1 \kappa_2 \Big(1 + \frac{\lambda_1}{2\kappa_1} \Big) 
\Big(1 + \frac{\lambda_2}{2\kappa_2} \Big)
\right.
\nonumber \\ && \hspace{40mm}
\left.
+ \kappa_1 \kappa_2 \Big(1 + \frac{\lambda_1}{2\kappa_2} \Big) 
\Big(1 + \frac{\lambda_2}{2\kappa_1} \Big)
+\kappa_2^2 \Big(1 + \frac{\lambda_1}{2\kappa_2} \Big) 
\Big(1 + \frac{\lambda_2}{2\kappa_2} \Big)
\right],
\label{I2-gen}
\end{eqnarray}
where $\Delta_i$'s are given by eq.~\eqref{Delta} after replacing
$\kappa \rightarrow \kappa_i$.

\subsubsection{Dirichlet limit}

For the case of the Dirichlet limit ($a \lambda_{1,2} \gg 1$)
the expression for $I^{(2)}(\kappa_1,\kappa_2)$ in eq.~\eqref{I2-gen}
takes on the relatively simple form
\begin{equation}
I^{(2)}_D(\kappa_1,\kappa_2)
= - \frac{\kappa_1}{\sinh \kappa_1 a} \frac{\kappa_2}{\sinh \kappa_2 a}
\label{I2D}
\end{equation} 
where the subscript $D$ stands for Dirichlet limit.
Using the above expression in eq.~\eqref{L2=I2} we have 
\begin{equation}
L^{(2)}_D(k_1,k_2) = \frac{1}{4\pi} 
\int_0^\infty \bar{\kappa} \,d\bar{\kappa}
\frac{\kappa_1}{\sinh \kappa_1 a} \frac{\kappa_2}{\sinh \kappa_2 a}.
\label{L2D}
\end{equation}
We describe the correspondence of the expression for the interaction
energy, in eq.~\eqref{DE12=L2},
in the Dirichlet limit, to the result in \cite{Emig:2003} in 
appendix \ref{app-emig}.

We shall digress a little to answer how the above result
is related to the Dirichlet Green's function.
We start by recalling the reduced Green's function 
corresponding to eq.~\eqref{g0zzp} in the Dirichlet limit,
which can be derived by taking the $\lambda_{1,2} \rightarrow \infty$ limit in 
$g_2^{(0)}(z,z^\prime;\kappa)$ in eq.~\eqref{g20zzp-sol} 
defined in region 2 ($a_1 < z,z^\prime < a_2$) in figure \ref{regions}, 
\begin{equation}
g_D^{(0)}(z,z^\prime;\kappa)
= \frac{\sinh \kappa (z_< - a_1) \sinh \kappa (a_2 - z_>)} 
       {\kappa \, \sinh \kappa a}
\label{g0zzp-D}
\end{equation}
where $z_<$ and $z_>$ stand for $\text{Min}(z,z^\prime)$
and $\text{Max}(z,z^\prime)$ respectively.
It is trivially verified that the above function satisfies the 
Dirichlet boundary conditions, 
$g^{(0)}_D(a_i,z^\prime;\kappa) = 0$ and $g^{(0)}_D(z,a_i;\kappa) = 0$.
Using eq.~\eqref{g0zzp-D} we can evaluate
\begin{equation}
\bigg[ \frac{d\;\;}{dz_<}\frac{d\;\;}{dz_>}
       \,g^{(0)}_D(z,z^\prime;\kappa) \bigg]_{z_<=a_1,z_>=a_2}
= - \frac{\kappa}{\sinh \kappa a}.
\end{equation}
Using the above result eq.~\eqref{I2D} can be expressed as 
the product of derivatives of two Dirichlet Green's functions.
It is also worth mentioning that the result in eq.~\eqref{I2D} can 
also be derived by exclusively using $g_2^{(0)}(z,z^\prime;\kappa)$ 
in eq.~\eqref{I2} after taking the limit 
$\lambda_{1,2} \rightarrow \infty$.
This is expected because the result in the Dirichlet limit should 
depend only on the quantities between the plates.
Thus, all the results related to the Dirichlet limit can be 
derived without relying on the averaging prescription for 
taking derivatives described in appendix \ref{ddz-green}.
However, for the more general case being considered here,
we require the averaging prescription to derive eq.~\eqref{I2-gen}.

\subsubsection{Weak coupling limit}

For the weak coupling limit ($a\lambda_{1,2} \ll 1$)
the expression for $I^{(2)}(\kappa_1,\kappa_2)$ in eq.~\eqref{I2-gen}
takes the very simple form
\begin{equation}
I^{(2)}_W(\kappa_1,\kappa_2)
= - \frac{\lambda_1}{2\kappa_1} \frac{\lambda_2}{2\kappa_2}
(\kappa_1 + \kappa_2)^2 \,e^{-a(\kappa_1 + \kappa_2)},
\label{I2W}
\end{equation}
where the subscript $W$ stands for the weak limit. 
We point out that in the case of weak coupling 
the averaging prescription for the evaluation of the $I$-kernels
was not necessary because only the free Green's functions come in.
Using the above expression in eq.~\eqref{L2=I2} we can evaluate
the corresponding $L$-kernel as
\begin{eqnarray}
L^{(2)}_W(k_1,k_2) 
&=& \frac{\lambda_1 \lambda_2}{16\pi} 
\int_0^\infty \bar{\kappa} \,d\bar{\kappa}
\,\frac{(\kappa_1 + \kappa_2)^2}{\kappa_1 \kappa_2} 
\,e^{-a(\kappa_1 + \kappa_2)}
= - \frac{\lambda_1 \lambda_2}{16\pi} 
\,\frac{\partial}{\partial a}
\bigg[ \frac{1}{a} e^{-a(|k_1|+|k_2|)} \bigg],
\label{L2W}
\end{eqnarray}
where in the evaluation of the integral we used the
change of variables $\kappa_1 + \kappa_2 = x$, 
and the corresponding relation
$\bar{\kappa} \,d\bar{\kappa} \,(\kappa_1 + \kappa_2) = \kappa_1\kappa_2 \,dx$.

\subsection{Sinusoidal corrugations: Leading order}

\begin{figure}
\begin{center}
\includegraphics[width=70mm]{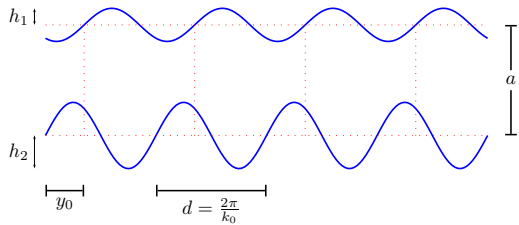}
\caption{Parallel plates with sinusoidal corrugations.}
\label{corru}
\end{center}
\end{figure}

For the particular case of sinusoidal corrugations, 
as described in figure \ref{corru}, we will have
\begin{subequations}
\begin{eqnarray}
h_1(y) &=& h_1 \sin [k_{0}(y + y_0)],
\\
h_2(y) &=& h_2 \sin [k_{0} y],
\end{eqnarray}\label{sin-p}\end{subequations}
where $k_0 = 2\pi/d$ is the wavenumber corresponding to 
the corrugation wavelength $d$. We get nonzero contributions,
to the leading order, only for the case when both plates have 
the same corrugation wavelength.
The Fourier transforms $\tilde{h}_i(k)$ for sinusoidal corrugations
get contributions in the form of delta functions
\begin{equation}
\tilde{h}_1(k) = h_1 \frac{2\pi}{2i}
\Big[ e^{ik_0y_0} \delta (k-k_0) - e^{-ik_0y_0} \delta (k+k_0) \Big].
\label{h-til}
\end{equation}
Using the above expression in eq.~\eqref{DE12=L2}, and after interpreting
$2\pi \delta(0) = L_y$ to be the infinite length in the $y$ direction,
we write 
\begin{equation}
\frac{E_{12}^{(2)}}{L_xL_y}
= \cos k_0 y_0 \, \frac{h_1 h_2}{4\pi}
\int_{-\infty}^{\infty} dk \,L^{(2)}(k,k_+)
= - \cos k_0 y_0 \, \frac{h_1 h_2}{16\pi^2} \int_{-\infty}^{\infty} dk 
\int_0^\infty \bar{\kappa} \,d\bar{\kappa} \,I^{(2)}(\kappa,\kappa_+),
\label{DE12=A2}
\end{equation}
where we have used the symmetry property in the $I^{(2)}$-kernel, 
noted in eq.~\eqref{I2sym}, and performed suitable rescaling 
in the integration variables.
We have used the notations $k_\pm=k\pm k_0$, and
$\kappa_\pm^2 = \bar{\kappa}^2 + k_\pm^2$.
The $I^{(2)}$-kernel in the above expression 
is provided by eq.~\eqref{I2-gen}.

\subsubsection{Dirichlet limit}

\begin{figure}
\begin{center}
\includegraphics[width=80mm]{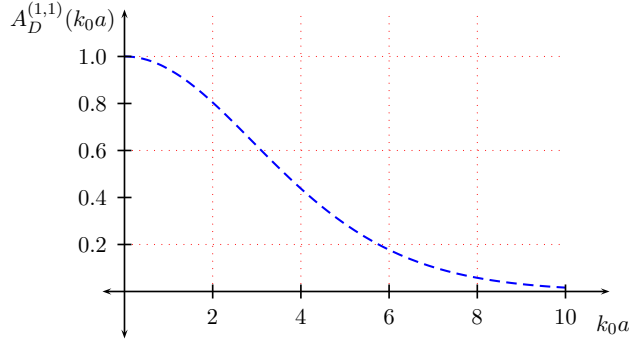}
\caption{Dirichlet limit: Plot of $A^{(1,1)}_D(k_0a)$ versus $k_0a$. }
\label{A11-D-t0-fig}
\end{center}
\end{figure}

In the Dirichlet limit, 
where the $I^{(2)}$-kernel is given by eq.~\eqref{I2D},
the interaction energy in eq.~\eqref{DE12=A2} can be expressed in the form
\begin{equation}
\frac{E_{12}^{(2)}}{L_xL_y}
= \cos (k_0 y_0) \,\frac{\pi^2}{240\,a^3} \,\frac{h_1}{a}\frac{h_2}{a}
\, A^{(1,1)}_D(k_0a),
\label{E12-2-p}
\end{equation}
where the function $A^{(1,1)}_D(k_0a)$ is normalized such that 
$A^{(1,1)}_D(0) = 1$, which becomes a convenience when we compare the 
results with those obtained by the proximity force approximation.
The expression for $A^{(1,1)}_D(k_0a)$ is
\begin{equation}
A^{(1,1)}_D(k_0a)
= \frac{60\,a^5}{\pi^3} \int_{-\infty}^{\infty}dk L^{(2)}_D(k,k_+)
= - \frac{1}{4\pi} \frac{60\,a^5}{\pi^3} \int_{-\infty}^{\infty}dk
\int_0^\infty \bar{\kappa}\,d\bar{\kappa} \, I^{(2)}_D(\kappa,\kappa_+),
\end{equation}
where the $I^{(2)}_D$-kernel, and the corresponding $L^{(2)}_D$-kernel,
in the Dirichlet limit were given in eq.~\eqref{I2D} and eq.~\eqref{L2D}
respectively.
Using the notations, $s^2 = \bar{s}^2 + t^2$,
$s^2_\pm = \bar{s}^2 + (t \pm t_0)^2$, and $t_0=k_0a$, we can write
\begin{equation}
A^{(1,1)}_D(t_0) = \frac{15}{\pi^4} \int_0^\infty \bar{s} \,d\bar{s}
\int_{-\infty}^{\infty} dt \frac{s}{\sinh s} \frac{s_+}{\sinh s_+}.
\label{Bt0}
\end{equation}
This expression can be numerically evaluated and has been 
plotted in figure \ref{A11-D-t0-fig}. 
The value of the function at $t_0=0$
can be evaluated exactly by rewriting the integral 
in terms of spherical polar coordinates to yield
\begin{equation}
A^{(1,1)}_D(0) = \frac{30}{\pi^4}
\int_0^\infty s^2 \,ds \,\bigg[ \frac{s}{\sinh s} \bigg]^2 =1.
\end{equation}
It should be mentioned that the result corresponding to eq.~\eqref{Bt0}
for the electromagnetic case and the scalar Dirichlet case have
been evaluated exactly in~\cite{Emig:2003}.

The lateral Casimir force can be evaluated using 
eq.~\eqref{E12-2-p} in the definition of lateral force in
eq.~\eqref{latf} which yields
\begin{equation}
F_\text{Lat,D}^{(2)} 
= 2 \,k_0a \,\sin (k_0y_0) \,\left| F_\text{Cas,D}^{(0)} \right| 
\,\frac{h_1}{a} \frac{h_2}{a} \, A^{(1,1)}_D(k_0a),
\label{latf-2D}
\end{equation}
where $|F_\text{Cas,D}^{(0)}|$ 
is the magnitude of the normal scalar Casimir force
between two uncorrugated parallel Dirichlet plates given as
\begin{equation}
\frac{F_\text{Cas,D}^{(0)}}{L_xL_y}
= - \frac{\pi^2}{480} \frac{1}{a^4}.
\end{equation}

\subsubsection{Weak coupling limit}

\begin{figure}
\begin{center}
\includegraphics[width=80mm]{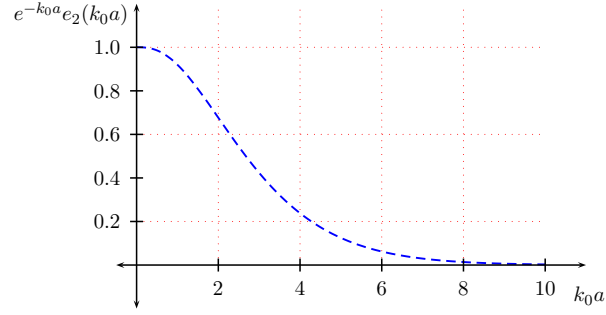}
\caption{Weak coupling limit:
Plot of $A^{(2)}_W(k_0a) = e^{-k_0a} e_2(k_0a)$ versus $k_0a$. }
\label{A2-W-t0-fig}
\end{center}
\end{figure}

Let us start by recalling the expression for the Casimir energy
and Casimir force between uncorrugated parallel plates in the 
weak limit:
\begin{eqnarray}
\frac{E_\text{Cas,W}^{(0)}}{L_xL_y}
= - \frac{\lambda_1 \lambda_2}{32\,\pi^2} \frac{1}{a}
\qquad \text{and} \qquad 
\frac{F_\text{Cas,W}^{(0)}}{L_xL_y}
= - \frac{\lambda_1 \lambda_2}{32\,\pi^2} \frac{1}{a^2}.
\end{eqnarray}
In the weak limit we have seen that it is possible to evaluate 
the $L$-kernel in eq.~\eqref{L2W} without any effort. Using this
in eq.~\eqref{DE12=A2} we have
\begin{equation}
\frac{E_{12,W}^{(2)}}{L_xL_y}
= - \cos (k_0y_0) \,\frac{\lambda_1 \lambda_2}{16\,\pi} 
\,\frac{h_1 h_2}{4\,\pi} \,\frac{\partial}{\partial a}
\bigg[ \frac{1}{a} \int_{-\infty}^{\infty} dk \,e^{-a(|k|+|k+k_0|)} \bigg],
\end{equation}
which after performing the integral can be written in the form
\begin{equation}
E_\text{12,W}^{(2)}
= \cos (k_0y_0) \,\left| E_\text{Cas,W}^{(0)} \right| 
\,\frac{h_1}{a} \frac{h_2}{a} \, A^{(2)}_W(k_0a),
\end{equation}
where we have introduced the function
\begin{equation}
A^{(2)}_W(t_0) = \frac{t_0^3}{2} \,\frac{\partial^2}{\partial t_0^2}
\bigg[ \frac{1}{t_0} \,e^{-t_0} \bigg]
= e^{-t_0} \sum_{m=0}^2 \, \frac{t_0^m}{m!} = \frac{e_2(t_0)}{e^{t_0}},
\end{equation}
where $e_n(x) = \sum_{m=0}^n \frac{x^m}{m!}$ is the the truncated 
exponential function which approximates to unity for $x\ll n$.
$A^{(2)}_W(t_0)$ has been plotted in figure \ref{A2-W-t0-fig}.
Using eq.~\eqref{latf} we evaluate the lateral Casimir force in 
this perturbation order as 
\begin{equation}
F_\text{Lat,W}^{(2)}
= k_0a \,\sin (k_0y_0) \,\left| F_\text{Cas,W}^{(0)} \right| 
\,\frac{h_1}{a} \frac{h_2}{a} \, A^{(2)}_W(k_0a).
\label{latf-2W}
\end{equation}

\section{Next-to-leading order contribution}

We start by collecting the terms in eq.~\eqref{formal-series}
in the form
\begin{equation}
E_{12} = E_{12}^{(2)} + E_{12}^{(3)} + E_{12}^{(4)} + {\cal O} (h/a)^5.
\end{equation}
To economize typographical space and for bookkeeping purposes
we introduce the notation
\begin{equation}
[G^{(0)}\Delta V]^{(m_1,m_2,\ldots,m_N)}_{(i_1,i_2,\ldots,i_N)}
= \text{Tr}
\Big[ G^{(0)} \Delta V_{i_1}^{(m_1)} G^{(0)} \Delta V_{i_2}^{(m_2)}
  \cdots G^{(0)} \Delta V_{i_N}^{(m_N)} \Big],
\label{GV-not}
\end{equation}
where the superscripts $m_n$ denote the $m_n$-th order contribution 
in the power series expansion of the potential as defined in 
eq.~\eqref{pot-series}. The subscripts $i_n$ identifies 
the potential that is contributing. In the case under consideration
we have only two potentials coupling and thus $i_n=1,2$ for any $n$.
As an illustrative example we have
\begin{equation}
[G^{(0)}\Delta V]^{(2,1,1)}_{(1,1,2)}
= \text{Tr}
\Big[ G^{(0)} \Delta V_1^{(2)} G^{(0)} \Delta V_1^{(1)}
  G^{(0)} \Delta V_2^{(1)} \Big].
\end{equation}
Using the above notation we can write
\begin{eqnarray}
E_{12}^{(3)}
&=& \frac{i}{2\tau}
\Big\{ [G^{(0)} \Delta V]^{(2,1)}_{(1,2)} + [G^{(0)} \Delta V]^{(1,2)}_{(1,2)}
     - [G^{(0)} \Delta V]^{(1,1,1)}_{(1,1,2)} 
     - [G^{(0)} \Delta V]^{(1,1,1)}_{(1,2,2)} \Big\}.
\end{eqnarray}

We argue that for the case of sinusoidal corrugations of the same 
wavelength on both plates the third order perturbation does not contribute. 
(Contributions from third order are nonzero if the wavelength 
of the corrugations of one plate is double that of the other plate.)
Since the plates under consideration are infinite in 
extent, the interaction energy is independent of translations in 
the $y$ direction. Let us now make a translation of the amount
$k_0 y = \pi$ in the potentials in eq.~\eqref{sin-p}. 
This amounts to replacing $h_i \rightarrow - h_i$.
Thus invariance under translation requires that the total power
of $h_i$'s in the result should be even. This rules out
the third order perturbation from contributing.

\subsection{Fourth order perturbation in Casimir energy due to corrugations}

Starting from eq.~\eqref{formal-series} and keeping terms contributing 
to the fourth order we have 
\begin{equation}
E_{12}^{(4)} = E_{12}^{(2,2)} + E_{12}^{(3,1)} + E_{12}^{(1,3)},
\label{DE12-4}
\end{equation}
where the superscripts represent the powers of $h_i$'s. For example,
the superscript $(3,1)$ represents terms involving $h_1^3 h_2$. 
All terms except one gets contribution from $m=1$ in the logarithm 
expansion in eq.~\eqref{formal-series}. Since the $m=2$ term is
distinct in structure we further separate it out by defining
\begin{equation}
E_{12}^{(2,2)} = E_{12}^{(2,2)A} + E_{12}^{(2,2)B},
\label{DE12-4AB}
\end{equation}
where the term involving the superscript $A$ is the contribution from 
$m=2$ in \eqref{formal-series}. Using the notation in eq.~\eqref{GV-not} 
we can thus collect the terms contributing
to eq.~\eqref{DE12-4} and eq.~\eqref{DE12-4AB} as
\begin{subequations}
\begin{eqnarray}
E_{12}^{(2,2)A}
&=& \frac{i}{2\tau} 
\Big\{ \frac{1}{2} \,[G^{(0)}\Delta V]^{(1,1,1,1)}_{(1,2,1,2)} \Big\},
\\
E_{12}^{(2,2)B}
&=& \frac{i}{2\tau}
\Big\{ [G^{(0)}\Delta V]^{(1,1,1,1)}_{(1,1,2,2)} 
      - [G^{(0)}\Delta V]^{(1,1,2)}_{(1,1,2)} 
      - [G^{(0)}\Delta V]^{(2,1,1)}_{(1,2,2)} 
      + [G^{(0)}\Delta V]^{(2,2)}_{(1,2)} \Big\},
\\
E_{12}^{(3,1)}
&=& \frac{i}{2\tau}
\Big\{ [G^{(0)}\Delta V]^{(1,1,1,1)}_{(1,1,1,2)}
      - [G^{(0)}\Delta V]^{(1,2,1)}_{(1,1,2)}
      - [G^{(0)}\Delta V]^{(2,1,1)}_{(1,1,2)}
      + [G^{(0)}\Delta V]^{(3,1)}_{(1,2)} \Big\},
\\
E_{12}^{(1,3)}
&=& \frac{i}{2\tau}
\Big\{ [G^{(0)}\Delta V]^{(1,1,1,1)}_{(1,2,2,2)}
      - [G^{(0)}\Delta V]^{(1,2,1)}_{(1,2,2)}
      - [G^{(0)}\Delta V]^{(1,1,2)}_{(1,2,2)}
      + [G^{(0)}\Delta V]^{(1,3)}_{(1,2)} \Big\}.
\end{eqnarray}\label{DE=GV-not}\end{subequations}
In terms of the reduced Green's function defined in 
eq.~\eqref{G0zzp} and eq.~\eqref{g0zzp}, and the Fourier transform 
of the corrugations in eq.~\eqref{hk=hy}, we can write
each of the terms in eq.~\eqref{DE=GV-not} in terms of
a corresponding $L$-kernel. In terms of the notation 
introduced in eq.~\eqref{GV-not} these will read as
\begin{equation}
(-1)^N \frac{i}{2\tau} 
\,[G^{(0)}\Delta V]^{(m_1,\ldots,m_N)}_{(i_1,\ldots,i_N)}
= L_x \int \frac{dk_1}{2\pi} \cdots \frac{dk_N}{2\pi}
\,\tilde{h}_{i_1}^{m_1}(k_1-k_2) \cdots \tilde{h}_{i_N}^{m_N}(k_N-k_1) 
\,L^{(m_1,\ldots,m_N)}_{(i_1,\ldots,i_N)}(k_1,\ldots,k_N),
\end{equation}
where implicitly we have interpreted the powers of the 
Fourier transformed corrugations as
\begin{equation}
\tilde{h}_{i_1}^{m_1}(k_1-k_2)
= \int \frac{dq_1}{2\pi} \cdots \frac{dq_{m_1-1}}{2\pi}
\,\tilde{h}_{i_1}(k_1-q_1) \tilde{h}_{i_1}(q_1-q_2) 
\cdots \tilde{h}_{i_1}(q_{m_1-1}-k_2). 
\end{equation}
Observe that each term contributing to the $M$-th order 
($M=m_1+m_2+\ldots+m_N$) has exactly $M$ $h$'s.
Also, the total number of variables inside the $L$-kernel is $N$ 
which is in general less than or equal to $M$.
It might be appropriate to call them $N$-point kernels.
Proceeding further, in the spirit of the second order calculation
in eq.~\eqref{L2=I2}, we introduce the corresponding $I$-kernels 
for the $L$-kernels as
\begin{equation}
L^{(m_1,\ldots,m_N)}_{(i_1,\ldots,i_N)}(k_1,\ldots,k_N)
= - \frac{1}{4\pi} \int_0^\infty \bar{\kappa} \,d \bar{\kappa}
\,\,I^{(m_1,\ldots,m_N)}_{(i_1,\ldots,i_N)}(\kappa_1,\ldots,\kappa_N),
\label{Lmi=Imi}
\end{equation}
where the $I$-kernels are expressed in terms of 
derivatives operating on the reduced Green's function
defined in eq.~\eqref{g0zzp} as
\begin{equation}
I^{(m_1,\ldots,m_N)}_{(i_1,\ldots,i_N)}(\kappa_1,\ldots,\kappa_N)
= (-1)^N \,\frac{\lambda_{i_1}}{m_1!} \cdots \frac{\lambda_{i_N}}{m_N!}
\frac{\partial^{m_1}}{\partial z_1^{m_1}}
       \cdots \frac{\partial^{m_N}}{\partial z_N^{m_N}}
\Big[g^{(0)}(z_N,z_1;\kappa_1) \cdots g^{(0)}(z_{N-1},z_N;\kappa_N)
\Big] \bigg|_{z_n = a_{i_n}}.
\label{Imi}
\end{equation}
For clarification we illustrate the evaluation of a particular term 
which should also serve as an illustration of the notation. 
The term we consider is
\begin{eqnarray}
-\frac{i}{2\tau} \, [G^{(0)} \Delta V]^{(1,2,1)}_{(1,1,2)}
&=& L_x \int \frac{dk_1}{2\pi} \frac{dk_2}{2\pi} \frac{dk_3}{2\pi}
\,\tilde{h}_1(k_1-k_2) \tilde{h}_1^2(k_2-k_3) \tilde{h}_2(k_3-k_1) 
\,L^{(1,2,1)}_{(1,1,2)}(k_1,k_2,k_3)
\nonumber \\
&=& L_x \int \frac{dk_1}{2\pi} \frac{dk_2}{2\pi}
    \frac{dk_3}{2\pi} \frac{dk_4}{2\pi}
\,\tilde{h}_1(k_1-k_2) \,\tilde{h}_1(k_2-k_3) 
\tilde{h}_1(k_3-k_4) \,\tilde{h}_2(k_4-k_1) 
\,L^{(1,2,1)}_{(1,1,2)}(k_1,k_2,k_4).
\hspace{5mm}
\end{eqnarray}
Notice how the particular $k$ dependence in the $L$-kernel
is unambiguously specified. The corresponding $I$-kernel 
using eq.~\eqref{Lmi=Imi} and eq.~\eqref{Imi} is given as
\begin{equation}
I^{(1,2,1)}_{(1,1,2)}(\kappa_1,\kappa_2,\kappa_4)
= - \frac{\lambda_1^2 \lambda_2}{2}
\frac{\partial}{\partial z_1} \frac{\partial^2}{\partial z_2^2}
       \frac{\partial}{\partial z_3}
\Big[ g^{(0)}(z_3,z_1;\kappa_1) g^{(0)}(z_1,z_2;\kappa_2) 
g^{(0)}(z_2,z_3;\kappa_4)
\Big] \bigg|_{{\scriptstyle z_1=a_1, z_2=a_1, z_3=a_2}}.
\end{equation}
Using the reciprocal symmetry in the Green's functions we can 
learn the following symmetries in the $I$-kernels associated
with the terms in eq.~\eqref{DE=GV-not}:
\begin{subequations}
\begin{align}
I^{(1,1,1,1)}_{(1,2,1,2)}(\kappa_1,\kappa_2,\kappa_3,\kappa_4)
&=I^{(1,1,1,1)}_{(1,2,1,2)}(\kappa_2,\kappa_1,\kappa_4,\kappa_3),
&&& I^{(1,1,2)}_{(1,1,2)}(\kappa_1,\kappa_2,\kappa_3)
&= I^{(1,1,2)}_{(1,1,2)}(\kappa_3,\kappa_2,\kappa_1),
\\
I^{(1,1,1,1)}_{(1,1,2,2)}(\kappa_1,\kappa_2,\kappa_3,\kappa_4)
&=I^{(1,1,1,1)}_{(1,1,2,2)}(\kappa_3,\kappa_2,\kappa_1,\kappa_4),
&&& I^{(2,1,1)}_{(1,2,2)}(\kappa_1,\kappa_2,\kappa_3)
&= I^{(2,1,1)}_{(1,2,2)}(\kappa_2,\kappa_1,\kappa_3),
\\
I^{(1,1,1,1)}_{(1,2,2,2)}(\kappa_1,\kappa_2,\kappa_3,\kappa_4)
&=I^{(1,1,1,1)}_{(1,2,2,2)}(\kappa_2,\kappa_1,\kappa_4,\kappa_3),
&&& I^{(1,2,1)}_{(1,2,2)}(\kappa_1,\kappa_2,\kappa_3)
&= I^{(1,1,2)}_{(1,2,2)}(\kappa_2,\kappa_1,\kappa_3),
\\
I^{(1,1,1,1)}_{(1,1,1,2)}(\kappa_1,\kappa_2,\kappa_3,\kappa_4)
&=I^{(1,1,1,1)}_{(1,1,1,2)}(\kappa_4,\kappa_3,\kappa_2,\kappa_1),
&&& I^{(1,2,1)}_{(1,1,2)}(\kappa_1,\kappa_2,\kappa_3)
&= I^{(2,1,1)}_{(1,1,2)}(\kappa_3,\kappa_2,\kappa_1).
\end{align}\label{I4sym}\end{subequations}
We have not listed the 2-point $I$-kernels in the above list because
they have the symmetry of the kind in eq.~\eqref{I2sym}.
In fact any 2-point kernel will have the following symmetry:
\begin{equation}
I^{(m_1,m_2)}_{(1,2)}(\kappa_1,\kappa_2)
= I^{(m_1,m_2)}_{(1,2)}(\kappa_2,\kappa_1).
\end{equation}

In the above discussion involving very general notations 
we have in principle expressed
each term in eq.~\eqref{DE=GV-not}. Specific evaluation of each term 
involves the evaluation of the corresponding $I$-kernels which are given
in terms of the derivatives of the Green's functions.
The derivatives are evaluated using the prescription described 
in appendix \ref{ddz-green}.
We can thus collect the terms in eq.~\eqref{DE=GV-not} into four 
distinct $L$-kernels in the form
\begin{subequations}
\begin{eqnarray}
\frac{E_{12}^{(2,2)A}}{L_x}
&=& \frac{1}{2} \int \frac{dk_1}{2\pi} \frac{dk_2}{2\pi} 
\frac{dk_3}{2\pi} \frac{dk_4}{2\pi}
\,\tilde{h}_1(k_1-k_2) \tilde{h}_2(k_2-k_3) 
\tilde{h}_1(k_3-k_4) \tilde{h}_2(k_4-k_1)
\,L^{(2,2)A}(k_1,k_2,k_3,k_4),
\label{E22A}
\hspace{3mm}
\\
\frac{E_{12}^{(2,2)B}}{L_x}
&=& \int \frac{dk_1}{2\pi} \frac{dk_2}{2\pi}
  \frac{dk_3}{2\pi} \frac{dk_4}{2\pi}
\,\tilde{h}_1(k_1-k_2) \tilde{h}_1(k_2-k_3)
\tilde{h}_2(k_3-k_4) \tilde{h}_2(k_4-k_1)
\,L^{(2,2)B}(k_1,k_2,k_3,k_4),
\\
\frac{E_{12}^{(3,1)}}{L_x}
&=& \int \frac{dk_1}{2\pi} \frac{dk_2}{2\pi}
  \frac{dk_3}{2\pi} \frac{dk_4}{2\pi}
\,\tilde{h}_1(k_1-k_2) \tilde{h}_1(k_2-k_3)
\tilde{h}_1(k_3-k_4) \tilde{h}_2(k_4-k_1)
\,L^{(3,1)}(k_1,k_2,k_3,k_4),
\\
\frac{E_{12}^{(1,3)}}{L_x}
&=& \int \frac{dk_1}{2\pi} \frac{dk_2}{2\pi}
  \frac{dk_3}{2\pi} \frac{dk_4}{2\pi}
\,\tilde{h}_1(k_1-k_2) \tilde{h}_2(k_2-k_3)
\tilde{h}_2(k_3-k_4) \tilde{h}_2(k_4-k_1)
\,L^{(1,3)}(k_1,k_2,k_3,k_4),
\end{eqnarray}\label{DE12=L4}\end{subequations}
where it should be noted that different $L$'s 
combine with specific combination of $h$'s.
The factor of one-half in eq.~\eqref{E22A} can be traced back to the 
coefficient of the second term in the expansion of logarithm in
eq.~\eqref{formal-series}.
The respective kernels $L^{(m,n)}$ above are related to their corresponding
$I^{(m,n)}$ by eq.~\eqref{Lmi=Imi}, which are given by 
\begin{subequations}
\begin{eqnarray}
I^{(2,2)A}(\kappa_1,\kappa_2,\kappa_3,\kappa_4)
&=& I^{(1,1,1,1)}_{(1,2,1,2)}(\kappa_1,\kappa_2,\kappa_3,\kappa_4),
\\
I^{(2,2)B}(\kappa_1,\kappa_2,\kappa_3,\kappa_4)
&=& I^{(1,1,1,1)}_{(1,1,2,2)}(\kappa_1,\kappa_2,\kappa_3,\kappa_4)
+ I^{(1,1,2)}_{(1,1,2)}(\kappa_1,\kappa_2,\kappa_3)
+ I^{(2,1,1)}_{(1,2,2)}(\kappa_1,\kappa_3,\kappa_4)
+ I^{(2,2)}_{(1,2)}(\kappa_1,\kappa_3),
\\
I^{(3,1)}(\kappa_1,\kappa_2,\kappa_3,\kappa_4)
&=& I^{(1,1,1,1)}_{(1,1,1,2)}(\kappa_1,\kappa_2,\kappa_3,\kappa_4)
+ I^{(1,2,1)}_{(1,1,2)}(\kappa_1,\kappa_2,\kappa_4)
+ I^{(2,1,1)}_{(1,1,2)}(\kappa_1,\kappa_3,\kappa_4)
+ I^{(3,1)}_{(1,2)}(\kappa_1,\kappa_4),
\\
I^{(1,3)}(\kappa_1,\kappa_2,\kappa_3,\kappa_4)
&=& I^{(1,1,1,1)}_{(1,2,2,2)}(\kappa_1,\kappa_2,\kappa_3,\kappa_4)
+ I^{(1,2,1)}_{(1,2,2)}(\kappa_1,\kappa_2,\kappa_4)
+ I^{(1,1,2)}_{(1,2,2)}(\kappa_1,\kappa_2,\kappa_3)
+ I^{(1,3)}_{(1,2)}(\kappa_1,\kappa_2).
\end{eqnarray}\label{I4-cons}\end{subequations}
Proceeding in the spirit of section \ref{leading-order}
has brought us to the point of evaluation of the thirteen
$I$-kernels on the right hand side of eq.~\eqref{I4-cons}.
Using the prescription described in appendix \ref{ddz-green}
we have evaluated all the thirteen kernels and explicit expressions
analogous to eq.~\eqref{I2-gen} have been derived. This would have
involved  a lot more labor and bookkeeping if not for the 
facilitation achieved by the use of Mathematica. 
We shall not display the explicit expressions
here because they are too long. But, as in section \ref{leading-order},
these expressions simplify considerably in the Dirichlet limit 
and the weak coupling limit.

\subsubsection{Dirichlet limit}

Observe that the reduced Green's function defined by eq.~\eqref{g0zzp}
has a well defined Dirichlet limit. In the light of this observation
in conjunction with the expression for the $I$-kernels in eq.~\eqref{Imi}
taken at face value suggests that these kernels might not have a
well-defined finite Dirichlet limit. However, we evaluated the second order
contribution in eq.~\eqref{I2D}. In fact it can be verified that
all 2-point kernels have a well-defined finite Dirichlet limit. Further,
$I^{(1,1,1,1)}_{(1,2,1,2)}$ also has a well-defined finite Dirichlet 
limit. The rest of the nine $I$-kernels on the right hand side
of eq.~\eqref{I4-cons} do not have a finite Dirichlet limit.
But, the sums of the $I$-kernels listed in eq.~\eqref{I4-cons}
have finite Dirichlet limits. The higher powers in 
$\lambda_{1,2}$ in the numerator of each of these sums cancel identically
to give a well-defined limit as $\lambda_{1,2} \rightarrow \infty$. 
This seems to be a generic phenomena.
Here we list the $I$-kernels in eq.~\eqref{I4-cons} evaluated
in the Dirichlet limit:
\begin{subequations}
\begin{eqnarray}
I^{(2,2)A}_D(\kappa_1,\kappa_2,\kappa_3,\kappa_4) 
&=& \frac{\kappa_1}{\sinh \kappa_1 a} \frac{\kappa_2}{\sinh \kappa_2 a}
\frac{\kappa_3}{\sinh \kappa_3 a} \frac{\kappa_4}{\sinh \kappa_4 a},
\\
I^{(2,2)B}_D(\kappa_1,\kappa_2,\kappa_3,\kappa_4) 
&=& \frac{\kappa_1}{\sinh \kappa_1 a} \frac{\kappa_2}{\tanh \kappa_2 a}
\frac{\kappa_3}{\sinh \kappa_3 a} \frac{\kappa_4}{\tanh \kappa_4 a},
\\
I^{(3,1)}_D(\kappa_1,\kappa_2,\kappa_3,\kappa_4) 
&=& - \frac{\kappa_1}{\sinh \kappa_1 a} \frac{\kappa_4}{\sinh \kappa_4 a}
\frac{1}{4} 
\left[ 4 \frac{\kappa_2}{\tanh \kappa_2 a} \frac{\kappa_3}{\tanh \kappa_3 a} 
+ \frac{\kappa_1^2 + \kappa_4^2}{3} - \kappa_2^2 - \kappa_3^2 \right],
\\
I^{(1,3)}_D(\kappa_1,\kappa_2,\kappa_3,\kappa_4)
&=& -\frac{\kappa_1}{\sinh \kappa_1 a} \frac{\kappa_2}{\sinh \kappa_2 a}
\frac{1}{4}
\left[ 4 \frac{\kappa_3}{\tanh \kappa_3 a} \frac{\kappa_4}{\tanh \kappa_4 a}
+ \frac{\kappa_1^2 + \kappa_2^2}{3} - \kappa_3^2 - \kappa_4^2 \right].
\end{eqnarray}\label{I4D}\end{subequations}

\subsubsection{Weak coupling limit}

We can similarly evaluate the $I$-kernels in the weak limit by 
keeping terms to order $\lambda_1 \lambda_2$ in the general expressions
for the $I$-kernels. But, we find it instructive, and simpler, to 
start over from eq.~\eqref{DE=GV-not}. We observe that only the 
2-point kernels contribute in the weak limit. We further notice that
only the contributions from the free Green's function (which is obtained 
by switching off the couplings $\lambda_{1,2}$ in eq.~\eqref{g0zzp})
contribute in the evaluation of the weak limit. These observations
allows us to write the 2-point kernels in the weak limit as
\begin{eqnarray}
I^{(m_1,m_2)}_{(1,2),W}(\kappa_1,\kappa_2)
&=& \frac{\lambda_1}{m_1!} \frac{\lambda_2}{m_2!}
\left[ \frac{\partial^{m_1}}{\partial z_1^{m_1}}
       \frac{\partial^{m_2}}{\partial z_2^{m_2}}
\frac{e^{-\kappa_1 |z_2-z_1|}}{2\kappa_1}
\frac{e^{-\kappa_2 |z_1-z_2|}}{2\kappa_2}
\right]_{z_1 = a_1, z_2=a_2}
\nonumber \\
&=& 
\frac{(-1)^{m_1}}{m_1! \, m_2!} 
\, \frac{\lambda_1 \lambda_2}{2\kappa_1 2\kappa_2}
\, (\kappa_1 + \kappa_2)^{M} \, e^{-a (\kappa_1 + \kappa_2)},
\end{eqnarray}
where $M=m_1+m_2$. 
Note that the derivatives in the above expressions
are well defined because they are evaluated at a point 
where $z_2 > z_1$.
In particular using these in eq.~\eqref{I4-cons} with the observation
that only the 2-point kernels contribute in the weak limit we get
$I^{(2,2)A}_W(\kappa_1,\kappa_2,\kappa_3,\kappa_4) = 0$ and
\begin{subequations}
\begin{align}
I^{(2,2)B}_W(\kappa_1,\kappa_2,\kappa_3,\kappa_4)
= I^{(2,2)}_{(1,2),W}(\kappa_1,\kappa_3)
&= \frac{1}{2!\,2!} \,\frac{\lambda_1 \lambda_2}{2\kappa_1 2\kappa_3}
\, (\kappa_1 + \kappa_3)^{4} \, e^{-a (\kappa_1 + \kappa_3)},
\\
I^{(3,1)}_W(\kappa_1,\kappa_2,\kappa_3,\kappa_4)
= I^{(3,1)}_{(1,2),W}(\kappa_1,\kappa_4)
&= \frac{1}{3!\,1!} \,\frac{\lambda_1 \lambda_2}{2\kappa_1 2\kappa_4}
\, (\kappa_1 + \kappa_4)^{4} \, e^{-a (\kappa_1 + \kappa_4)},
\\
I^{(1,3)}_W(\kappa_1,\kappa_2,\kappa_3,\kappa_4)
= I^{(1,3)}_{(1,2),W}(\kappa_1,\kappa_2)
&= \frac{1}{1!\,3!} \,\frac{\lambda_1 \lambda_2}{2\kappa_1 2\kappa_2}
\, (\kappa_1 + \kappa_2)^{4} \, e^{-a (\kappa_1 + \kappa_2)},
\end{align}\label{I4W}\end{subequations}
where notice how they differ in the particular dependence on the variables.
We again (see comments after eq.~\eqref{I2W}) point out that
the averaging prescription for the evaluation of the $I$-kernels
was not necessary to arrive at the above expressions.

Using the change of variables introduced in evaluating eq.~\eqref{L2W}
we can similarly evaluate the $L$-kernels using eq.~\eqref{Lmi=Imi} 
in the weak limit as
\begin{equation}
L^{(m_1,m_2)}_{(1,2),W}(k_1,k_2)
= \frac{(-1)^{m_2}}{m_1! \, m_2!}
\, \frac{\lambda_1 \lambda_2}{16\,\pi}
\, \left( \frac{\partial}{\partial a} \right)^{M-1}
\bigg[ \frac{1}{a} \,e^{-a(|k_1|+|k_2|)} \bigg],
\label{L2W-gen}
\end{equation}
which reproduces the result in eq.~\eqref{L2W} for $M=2$.

\subsection{Sinusoidal corrugations: Next-to-leading order}

We shall now specialize to the particular case of sinusoidal 
corrugations described by eq.~\eqref{sin-p}. Using the Fourier
transforms in eq.~\eqref{h-til} all but one of the $k$ integrals 
in eq.~\eqref{DE12=L4} can be immediately performed.
The symmetries listed in eq.~\eqref{I4sym} lead to simplifications
in the expressions. As in section \ref{leading-order} the 
expressions boil down to two integrals in variables $k$ and 
$\bar{\kappa}$. In particular eq.~\eqref{DE12=L4} takes the form
\begin{subequations}
\begin{eqnarray}
\frac{E^{(2,2)}_{12}}{L_xL_y}
&=& - \cos (2k_0y_0) \, \frac{h_1^2 h_2^2}{64\pi^2} 
\int_{-\infty}^{\infty} dk\int_0^\infty \bar{\kappa} \,d\bar{\kappa}
\Big[ \,\frac{1}{2} \,I^{(2,2)A}(\kappa,\kappa_+,\kappa,\kappa_+)
+ I^{(2,2)B}(\kappa_-,\kappa,\kappa_+,\kappa) \Big],
\\
\frac{E^{(3,1)}_{12}}{L_xL_y}
&=& - \cos (k_0y_0) \, \frac{h_1^3 h_2}{64\pi^2}
\int_{-\infty}^{\infty} dk\int_0^\infty \bar{\kappa} \,d\bar{\kappa}
\Big[ I^{(3,1)}(\kappa,\kappa_+,\kappa,\kappa_+)
+ I^{(3,1)}(\kappa,\kappa_+,\kappa,\kappa_-)
+ I^{(3,1)}(\kappa,\kappa_-,\kappa,\kappa_+) \Big],
\hspace{8mm}
\\
\frac{E^{(1,3)}_{12}}{L_xL_y}
&=& - \cos (k_0y_0) \, \frac{h_1 h_2^3}{64\pi^2}
\int_{-\infty}^{\infty} dk\int_0^\infty \bar{\kappa} \,d\bar{\kappa}
\Big[ I^{(1,3)}(\kappa,\kappa_+,\kappa,\kappa_+)
+ I^{(1,3)}(\kappa,\kappa_+,\kappa,\kappa_-)
+ I^{(1,3)}(\kappa,\kappa_-,\kappa,\kappa_+) \Big],
\end{eqnarray}\label{DE12=I4}\end{subequations}
where the $I$-kernels are given from eq.~\eqref{I4-cons}.
We use the notations $k_\pm$ and $\kappa_\pm$ introduced after
eq.~\eqref{DE12=A2}. We note the factor of $2$ in the argument
of cosine function of the first term above is a mark of
the fourth order in perturbation theory.
It should be mentioned here that in the above expressions we have 
omitted finite terms not having any dependence in the lateral shift
variable $y_0$. These terms do not contribute to the lateral force.

\subsubsection{Dirichlet limit}

We presented the expressions for the $I$-kernels in the Dirichlet limit
in eq.~\eqref{I4D}. Using these in eq.~\eqref{DE12=I4} and then 
adding the contributions from the three terms in eq.~\eqref{DE12-4}
we can write the total contribution to interaction energy due to 
the presence of corrugations as
\begin{equation}
\frac{E_{12}^{(4)}}{L_xL_y}
= \frac{\pi^2}{240\,a^3} \frac{h_1}{a} \frac{h_2}{a} \frac{15}{4}
\left[ \,\cos (k_0y_0) 
\left\{ \frac{h_1^2}{a^2} \,A^{(3,1)}_D(k_0a)
+ \frac{h_2^2}{a^2} \,A^{(1,3)}_D(k_0a)
\right\}
- \cos (2k_0y_0) \,\frac{1}{2} \frac{h_1}{a} \frac{h_2}{a} \,A^{(2,2)}_D(k_0a)
\right], 
\end{equation}
where we have introduced the functions
\begin{subequations}
\begin{eqnarray}
A^{(3,1)}_D(t_0) = A^{(1,3)}_D(t_0)
&=& \frac{1}{2\pi^4} \int_0^\infty \bar{s}\,d\bar{s} 
\int_{-\infty}^{\infty} dt 
\frac{s}{\sinh s} \frac{s_+}{\sinh s_+}
\left[ 4 \frac{s}{\tanh s} \frac{s_-}{\tanh s_-}
+ 2 \frac{s}{\tanh s} \frac{s_+}{\tanh s_+} - s^2 - s_-^2 \right],
\hspace{7mm}
\\
A^{(2,2)}_D(t_0) &=& \frac{1}{\pi^4} \int_0^\infty \bar{s}\,d\bar{s} 
\int_{-\infty}^{\infty} dt
\left[ \frac{s^2}{\sinh^2 s} \frac{s_-^2}{\sinh^2 s_-}
+ 2 \frac{s^2}{\tanh^2 s} \frac{s_+}{\sinh s_+} \frac{s_-}{\sinh s_-}
\right].
\end{eqnarray}\end{subequations}
\begin{figure}
\begin{tabular}{cc}
\includegraphics[width=80mm]{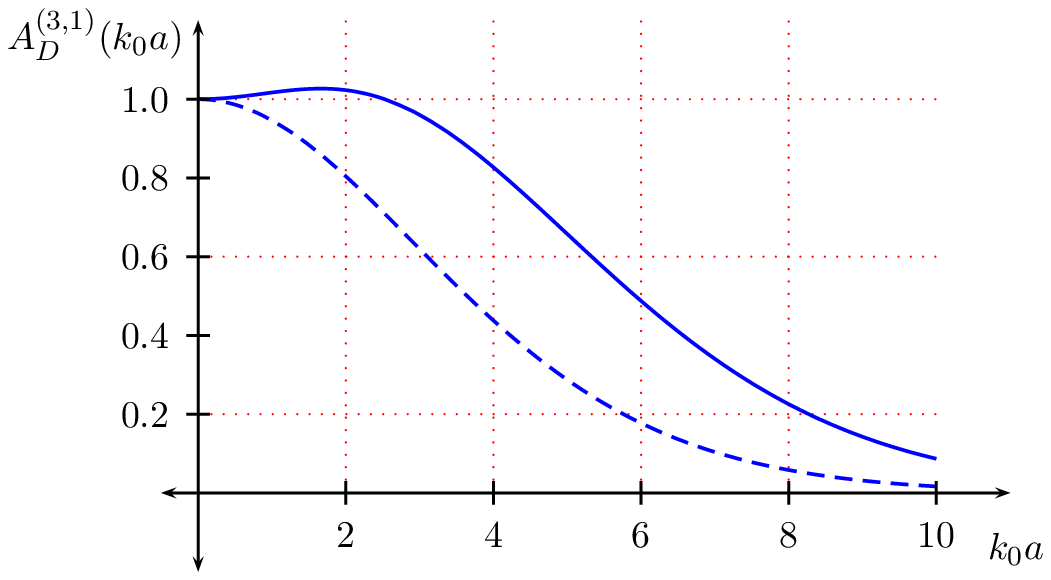} &
\includegraphics[width=80mm]{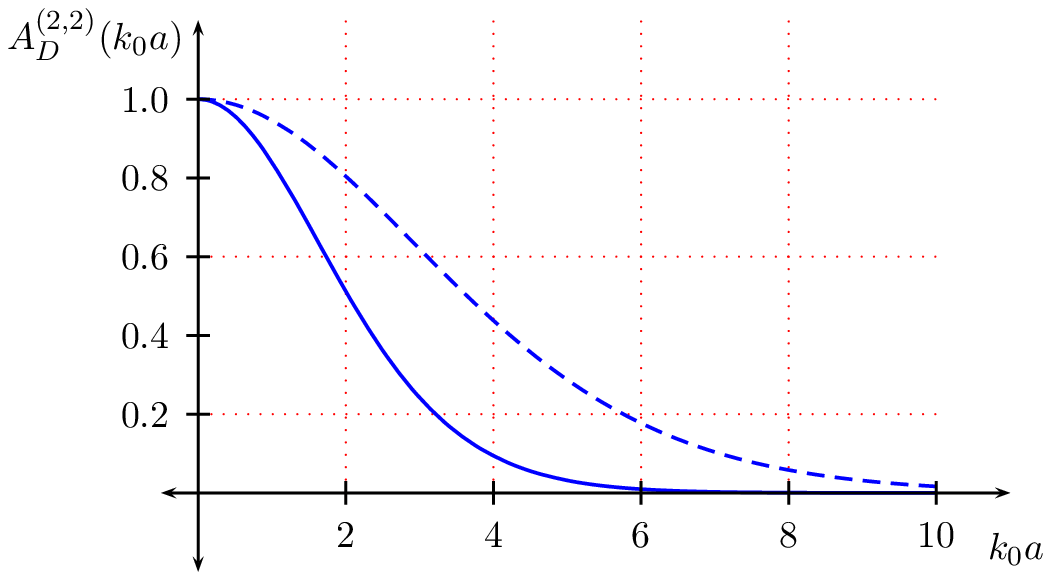}
\end{tabular}
\caption{
Dirichlet limit:
Plot of $A^{(3,1)}_D(k_0a)$ and $A^{(2,2)}_D(k_0a)$
versus $k_0a$. The dashed curve represents $A^{(1,1)}_D(k_0a)$
which is plotted here for reference. }
\label{A3122-W-t0-fig}
\end{figure}
These functions are plotted in figure~\ref{A3122-W-t0-fig}.
As in section \ref{leading-order} the above functions are suitably
normalized such that they evaluate to unity when evaluated at $t_0=0$.
Verification of this statement involves the following integrals:
\begin{subequations}
\begin{eqnarray}
A^{(3,1)}_D(0)
&=& \frac{2}{\pi^4} \int_0^\infty s^2\,ds
\frac{s^2}{\sinh^2 s} \left[ 3 \frac{s^2}{\tanh^2 s} - s^2 \right]
= \Big(1 + \frac{\pi^2}{21} \Big) -\frac{\pi^2}{21} = 1,
\\
A^{(2,2)}_D(0)
&=& \frac{2}{\pi^4} \int_0^\infty s^2\,ds
\left[ \frac{s^4}{\sinh^4 s} + 2 \frac{s^2}{\tanh^2 s} \frac{s^2}{\sinh^2 s}
\right]
= \Big(\frac{1}{3} + \frac{2\,\pi^2}{63} \Big) 
 + \Big(\frac{2}{3} - \frac{2\,\pi^2}{63} \Big) =1.
\end{eqnarray}\end{subequations}

Therefore, 
the next-to-leading-order contribution to the lateral Casimir force 
in the Dirichlet limit reads
\begin{equation}
F_\text{Lat,D}^{(4)}
= 2 \,k_0a \, \sin (k_0y_0) \,\left|F_\text{Cas,D}^{(0)} \right| 
\,\frac{h_1}{a} \frac{h_2}{a} \,\frac{15}{4}
\left[ \left( \frac{h_1^2}{a^2} + \frac{h_2^2}{a^2} \right) A^{(3,1)}_D(k_0a)
- 2 \cos (k_0y_0) \,\frac{h_1}{a} \frac{h_2}{a} \,A^{(2,2)}_D(k_0a) \right].
\label{latf-4D}
\end{equation}

\subsubsection{Weak coupling limit}

We have already evaluated the $L$-kernel for the weak coupling case
in eq~\eqref{L2W-gen},
so we can immediately evaluate the interaction energy.
Alternatively, we can use
the expressions for the $I$-kernels in the weak coupling limit
(see eq.~\eqref{I4W}) into eq.~\eqref{DE12=I4} 
to evaluate the integrals exactly in the spirit of section
\ref{leading-order} and the answer can be expressed as derivatives.
Using these in
\eqref{DE12-4} we can write the total contribution to the interaction
energy in the presence of corrugations in the weak limit to be
\begin{equation}
E_{12,W}^{(4)}
= \left| E^{(0)}_\text{Cas,W} \right| \frac{h_1}{a} \frac{h_2}{a} \frac{3}{2}
\left[ \cos (k_0y_0) 
\left( \frac{h_1^2}{a^2} + \frac{h_2^2}{a^2} \right) A^{(4)}_W(k_0a)
- \cos (2k_0y_0) \frac{1}{2} \frac{h_1}{a} \frac{h_2}{a} A^{(4)}_W(2k_0a)
\right],
\label{DE12-4W}
\end{equation}
where in contrast to the result in the Dirichlet limit we observe
the appearance of the same function, though with different arguments,
as coefficients. This function has been suitably defined as
\begin{equation}
A^{(4)}_W(t_0) = \frac{t_0^5}{4!} \frac{\partial^4}{\partial t_0^4}
\bigg[ \frac{1}{t_0} \, e^{-t_0} \bigg]
=e^{-t_0} \sum_{m=0}^4 \, \frac{t_0^m}{m!} = \frac{e_4(t_0)}{e^{t_0}},
\end{equation}
\begin{figure}
\begin{tabular}{cc}
\includegraphics[width=80mm]{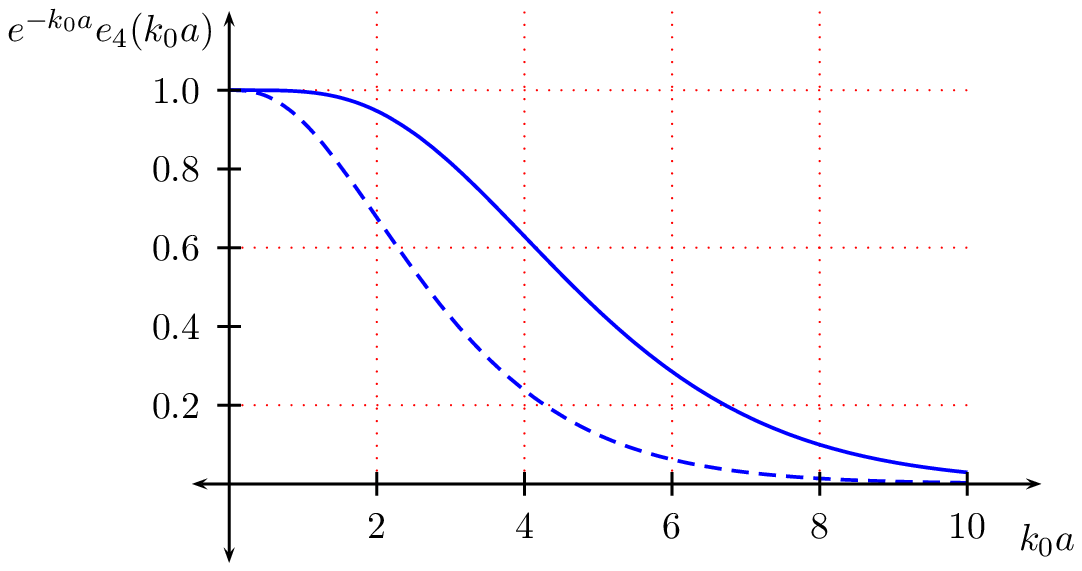} &
\includegraphics[width=80mm]{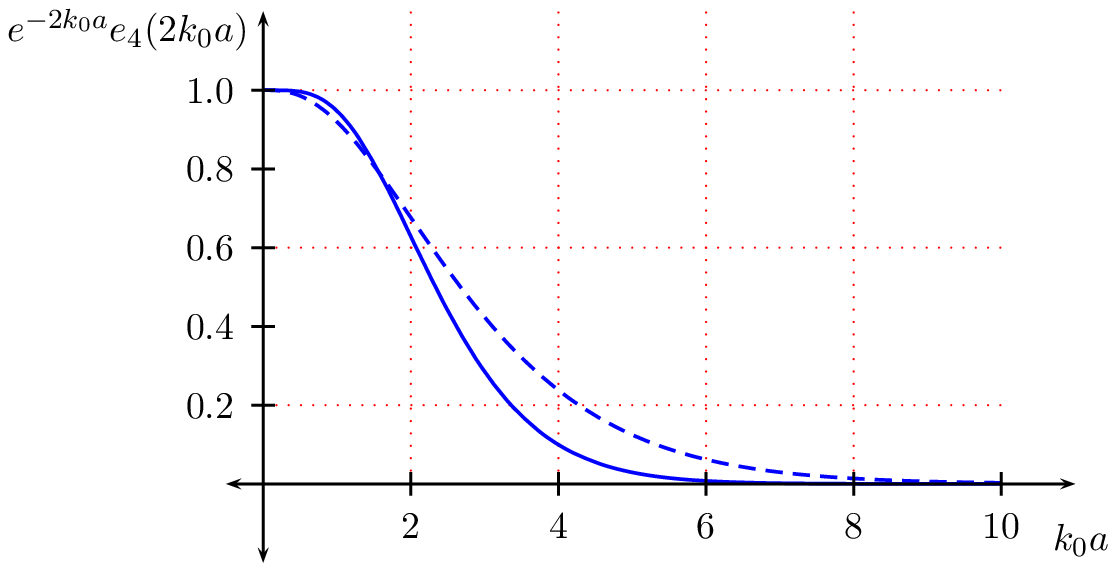}
\end{tabular}
\caption{
Weak coupling limit:
Plot of $A^{(4)}_W(k_0a)$ and $A^{(4)}_W(2k_0a)$
versus $k_0a$. The dashed curve represents $A^{(2)}_W(k_0a)$
which is plotted here for reference. }
\label{A4-W-t0-fig}
\end{figure}
so that it equals unity at $t_0=0$. 
These are plotted in figure \ref{A4-W-t0-fig}.
Using eq.~\eqref{DE12-4W} in eq.~\eqref{latf} we find that the fourth order
contribution to the lateral Casimir force in the weak coupling limit
equals
\begin{equation}
F_\text{Lat,W}^{(4)}
= k_0a \, \sin (k_0y_0) \,\left|F_\text{Cas,W}^{(0)} \right|
\,\frac{h_1}{a} \frac{h_2}{a} \,\frac{3}{2}
\left[ \left( \frac{h_1^2}{a^2} + \frac{h_2^2}{a^2} \right) A^{(4)}_W(k_0a)
- 2 \cos (k_0y_0) \,\frac{h_1}{a} \frac{h_2}{a} \,A^{(4)}_W(2k_0a) \right].
\label{latf-4W}
\end{equation}

An exact functional form for the lateral force in the weak coupling limit
in the second and fourth order in the perturbative expansion,
see eq.~\eqref{latf-2W} and eq.~\eqref{latf-4W},
allows us to investigate the approximation involved in the perturbative
analysis more closely. We observe that in each perturbative order the 
leading powers of $k_0a$ in the $A$-function cancels the $a$ dependence
in the prefactors associated with $h_i$.
Thus we conclude that the perturbation analysis
is valid for $k_0h \ll 1$.
The formal expansions performed in the intermediate steps further
restricts our analysis to be valid for $h_1 +h_2 < a$
for arbitrary offset.

\section{Proximity force approximation}

As a preliminary check for the results calculated for 
the lateral force we shall verify that in the proximity 
force limit ($k_0a \rightarrow 0$, keeping $h_i/a$ fixed,) 
our results in 
eqs.~\eqref{latf-2D}, \eqref{latf-2W}, \eqref{latf-4D}, and \eqref{latf-4W},
reproduce the standard PFA result approximated to appropriate 
orders in $h_i/a$. 
For this purpose let us evaluate the lateral Casimir force
in the proximity force approximation for the sinusoidal corrugations
described by eq.~\eqref{sin-p}.
We begin by writing the distance between the corrugated plates as 
\begin{equation}
a(y) = a + h_2 \sin[k_0 y] - h_1 \sin [k_0(y + y_0)].
\end{equation}
For sufficiently small $a$ (in comparison to $d=2\pi \,k_0^{-1}$) 
we can treat the plates to be built out of small sections of length
$dy$ for which the energy is approximately that of the parallel plate 
geometry. 

\setcounter{subsubsection}{0}
\subsubsection{Dirichlet limit}

Using the expression for the Casimir force between parallel plates
in the Dirichlet limit we can thus write
\begin{equation}
dE^\text{PFA}_D(a(y)) = - L_x \,dy \,\frac{\pi^2}{1440} \frac{1}{a(y)^3},
\label{dEPFA-D}
\end{equation}
where we interpret $L_x \, dy$ to be the area of the small section
under consideration. Thus the total Casimir
energy in this approximation,
after interpreting $L_y = \lim_{N\rightarrow\infty} N d$,
will be
\begin{eqnarray}
\frac{E^\text{PFA}_D}{L_xL_y}
&=& - \frac{\pi^2}{1440} \,\frac{1}{a^3} \,\frac{1}{2\pi}
\int_{-\pi}^{\pi} d\theta \frac{1}{\left[1 + \frac{h_2}{a} \sin \theta 
         - \frac{h_1}{a} \sin (\theta+k_0y_0)\right]^3}
\nonumber \\
&=& - \frac{\pi^2}{1440} \,\frac{1}{a^3} \,\frac{1}{2\pi}
\int_{-\pi}^{\pi} d\theta \frac{1}{\left[1 - \frac{r}{a} \cos \theta \right]^3},
\label{EDPFA}
\end{eqnarray}
where we have used the substitutions:
$r \sin\alpha = h_2 - h_1 \cos k_0y_0$, $r \cos\alpha = h_1 \sin k_0y_0$,
and used the periodicity property of the function to eliminate $\alpha$.
We note that $r^2 = h_1^2 + h_2^2 - 2 h_1 h_2 \cos (k_0y_0)$.

As was emphasized in the calculations in the earlier 
sections we have been subtracting the energies that do not 
contribute to the torque (or the lateral force) by treating it
as a background. This has not been achieved in the above expression
in eq.~\eqref{EDPFA}
and thus should not be naively compared with our earlier results.
But, evaluation of the lateral force will let us compare the 
expressions. Thus, we calculate the lateral Casimir force in 
the proximity force approximation, using 
eq.~\eqref{EDPFA} in eq.~\eqref{latf}, to be
\begin{equation}
F_\text{Lat,D}^\text{PFA}
= 2 \,k_0a \,\sin (k_0y_0) \,\left| F_\text{Cas,D}^{(0)} \right| 
\,\frac{h_1}{a} \frac{h_2}{a} \, \frac{1}{4}
\left[ \frac{5}{(1-\frac{r^2}{a^2})^\frac{7}{2}} 
       - \frac{1}{(1-\frac{r^2}{a^2})^\frac{5}{2}} \right],
\quad \text{for} \quad |h_1| + |h_2| < a.
\label{latf-PFA-D}
\end{equation}
The basic integral involved in evaluating the above expression is
\begin{equation}
\frac{1}{2\pi} \int_{-\pi}^{\pi} d\theta
\, \frac{1}{(a - r \cos \theta)} = \frac{1}{\sqrt{a^2 - r^2}},
\quad \text{for} \quad r < a,
\label{a+rcos-int}
\end{equation}
which, after the substitution $z = \exp(i\theta)$, leads to a contour 
integral on the unit circle in the complex plane. Thus, using 
Cauchy's theorem only those roots of the denominator that are within
the unit circle contribute to the integral.

The PFA result in the leading order for the perturbation parameter is
obtained by setting $r=0$ in eq.~\eqref{latf-PFA-D}. This matches
exactly with eq.~\eqref{latf-2D} to the leading order in $k_0a$.
This verifies the proximity force limit of the leading order
perturbative result. The proximity force limit of the 
next-to-leading-order result in eq.~\eqref{latf-4D} is similarly 
verified by matching it with eq.~\eqref{latf-PFA-D} after 
evaluating eq.~\eqref{latf-PFA-D} to the fourth order in 
the perturbation parameter.

\subsubsection{Weak coupling limit}

In the weak coupling limit and in the proximity approximation the 
equation corresponding to eq.~\eqref{dEPFA-D} is
\begin{equation}
dE^\text{PFA}_W(a(y)) = - L_x \,dy \,\frac{\lambda_1 \lambda_2}{32\,\pi^2} 
\frac{1}{a(y)}
\label{dEPFA-W}
\end{equation}
and leads to, after extracting $L_y$,
and proceeding in the same manner as in eq.~\eqref{EDPFA},
\begin{eqnarray}
\frac{E^\text{PFA}_W}{L_xL_y}
&=& - \frac{\lambda_1 \lambda_2}{32\,\pi^2} \,\frac{1}{a} \,\frac{1}{2\pi}
\int_{-\pi}^{\pi} d\theta \frac{1}{(1 + \frac{r}{a} \cos \theta)}
= - \frac{\lambda_1 \lambda_2}{32\,\pi^2} \,\frac{1}{a} 
\frac{1}{(1 - \frac{r^2}{a^2})^\frac{1}{2}},
\quad \text{for} \quad |h_1| + |h_2| < a.
\end{eqnarray}
We calculate the lateral Casimir force in the weak coupling limit
in the proximity force approximation, using eq.~\eqref{latf}, to be
\begin{equation}
F^\text{PFA}_\text{Lat,W}
= k_0a \,\sin (k_0y_0) \,\left| F^{(0)}_\text{Cas,W} \right|
\,\frac{h_1}{a} \frac{h_2}{a} \, \frac{1}{(1 - \frac{r^2}{a^2})^\frac{3}{2}},
\quad \text{for} \quad |h_1| + |h_2| < a.
\label{latf-PFA-W}
\end{equation}
The proximity force limit of the leading- and next-to-leading-order
results in the weak coupling limit, eq.~\eqref{latf-2W} and
eq.~\eqref{latf-4W}, can be verified by keeping the appropriate
terms in eq.~\eqref{latf-PFA-W}.

\section{Analysis and discussion}

In this section we shall exclusively deal either in the Dirichlet 
limit or in the weak coupling limit. We shall further confine 
ourselves to the case $h_1 = h_2 = h$. Also, we shall find it convenient
to define the following dimensionless quantities 
\begin{subequations}
\begin{eqnarray}
{\cal F}_{D} \left( k_0a,\frac{h}{a},k_0y_0 \right)
&=& \frac{F_\text{Lat,D} \left(k_0a,\frac{h}{a},k_0y_0 \right)}
       {2 \, \left| F^{(0)}_\text{Cas,D} \right| 
        \,\frac{h^2}{a^2} \,k_0a \,\sin (k_0y_0)},
\\
{\cal F}_{W} \left( k_0a,\frac{h}{a},k_0y_0 \right)
&=& \frac{F_\text{Lat,W} \left(k_0a,\frac{h}{a},k_0y_0 \right)}
        {\left| F^{(0)}_\text{Cas,W} \right|
        \,\frac{h^2}{a^2} \,k_0a \,\sin (k_0y_0)}.
\end{eqnarray}\label{dless-force}\end{subequations}
We shall use superscripts $[m,n]$ on these dimensionless
quantities to denote the order in the power series expansion
to which the lateral force (not ${\cal F}$) have been calculated. 
In particular $m$ will signify the order in the 
parameter $k_0a$, and $n$ will denote the order in $h/a$.
The result in the proximity force approximation should be valid to the
first order in $k_0a$. Thus ${\cal F}^{[1,\infty]}$ will represent 
the PFA result in our notation. The perturbative results are complete
in the $k_0a$ dependence and thus will be denoted by 
${\cal F}^{[\infty,n]}$. We use square brackets to avoid the 
confusion it might lead to with the already introduced notation
to denote the powers in $h_i$'s in the superscripts of the $A$-functions.
In these notations our results can be summarized as,
(see eqs.~\eqref{latf-2D}, \eqref{latf-2W}, \eqref{latf-4D},
\eqref{latf-4W}, \eqref{latf-PFA-D}, and \eqref{latf-PFA-W},)
\begin{subequations}
\begin{align}
{\cal F}_{D}^{[\infty,4]}
&= A^{(1,1)}_D(k_0a) +\frac{15}{2} \frac{h^2}{a^2} 
\Big[ A^{(3,1)}_D(k_0a) - \cos (k_0y_0) \,A^{(2,2)}_D(k_0a) \Big],
&{\cal F}_{D}^{[1,\infty]}
&= \frac{1}{4}
\left[ \frac{5}{(1-\frac{r^2}{a^2})^\frac{7}{2}} 
-\frac{1}{(1-\frac{r^2}{a^2})^\frac{5}{2}} \right],
\\
{\cal F}_{W}^{[\infty,4]}
&= \frac{e_2(k_0a)}{e^{k_0a}} + 3 \frac{h^2}{a^2}
\bigg[ \frac{e_4(k_0a)}{e^{k_0a}} 
- \cos (k_0y_0) \,\frac{e_4(2k_0a)}{e^{2k_0a}} \bigg],
&{\cal F}_{W}^{[1,\infty]} &= \frac{1}{(1-\frac{r^2}{a^2})^\frac{3}{2}},
\end{align}\label{latf-frac}\end{subequations}
where we recall the definition of $r$ after eq.~\eqref{EDPFA}.
As we mentioned earlier these expressions match to the order $[1,4]$
both in the Dirichlet and weak limit.
Recall, as justified after eq.~\eqref{latf-4W}, that $k_0h$ is
the perturbative parameter. Thus,
we shall find it convenient to keep $k_0h$ fixed while discussing 
the perturbative results. This should be contrasted with the 
case of proximity force approximation where it is more 
convenient to keep the parameter $h/a$ fixed.
All of the results in eq.~\eqref{latf-frac} are valid for $2h <a$.

\setcounter{subsubsection}{0}
\subsubsection{Dirichlet limit}

\begin{figure}
\begin{tabular}{cc}
\includegraphics[width=80mm]{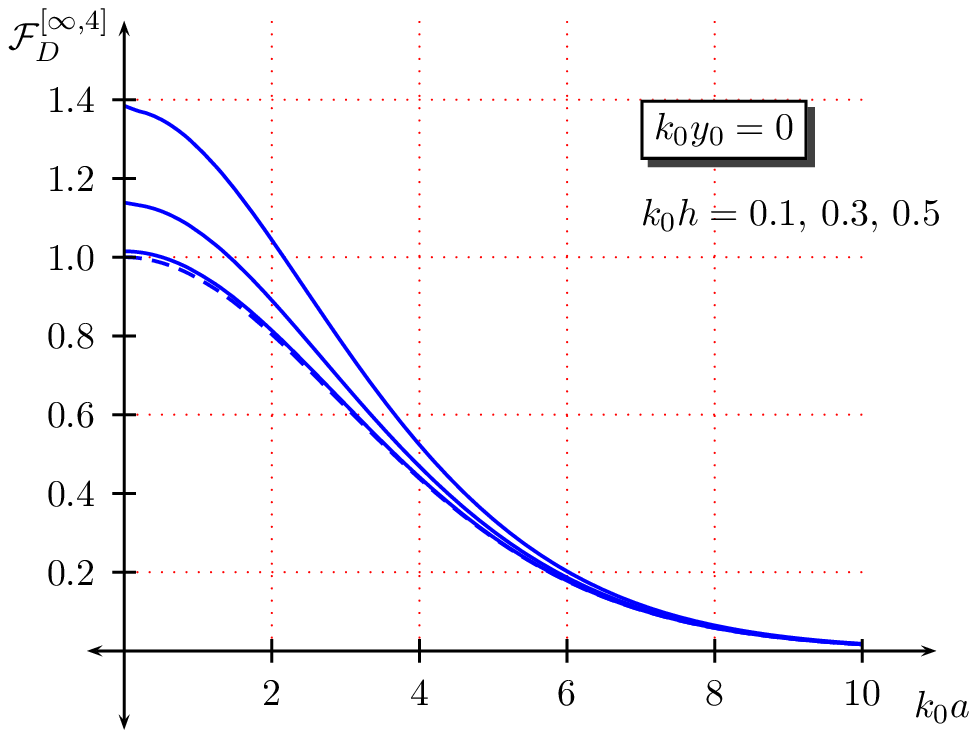} &
\includegraphics[width=80mm]{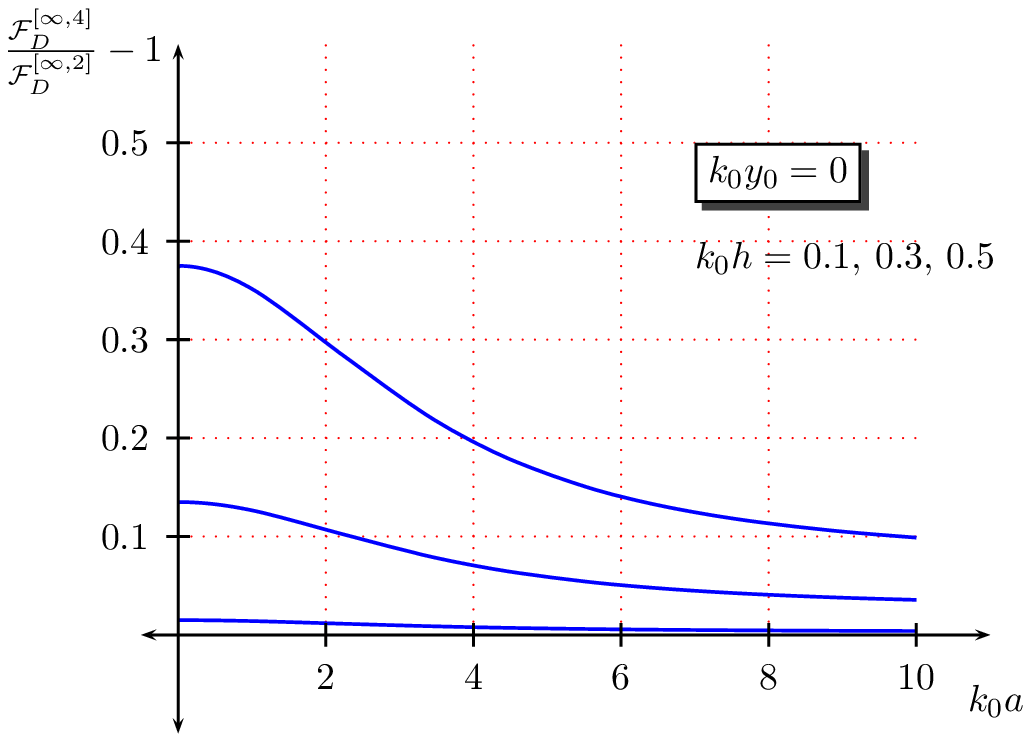}
\\
\includegraphics[width=80mm]{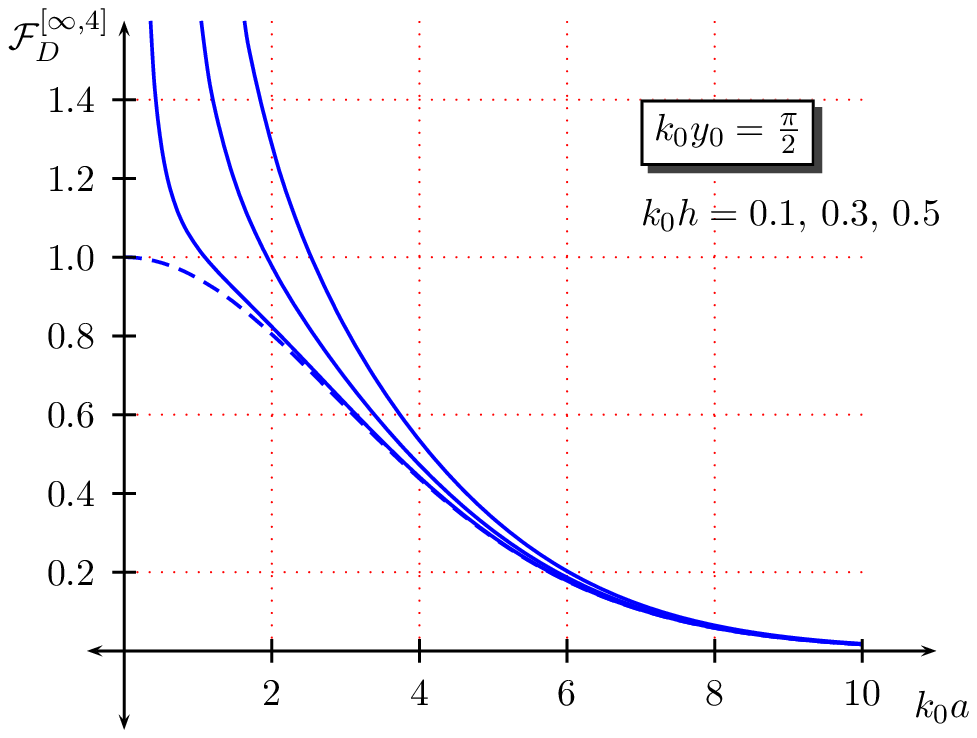} &
\includegraphics[width=80mm]{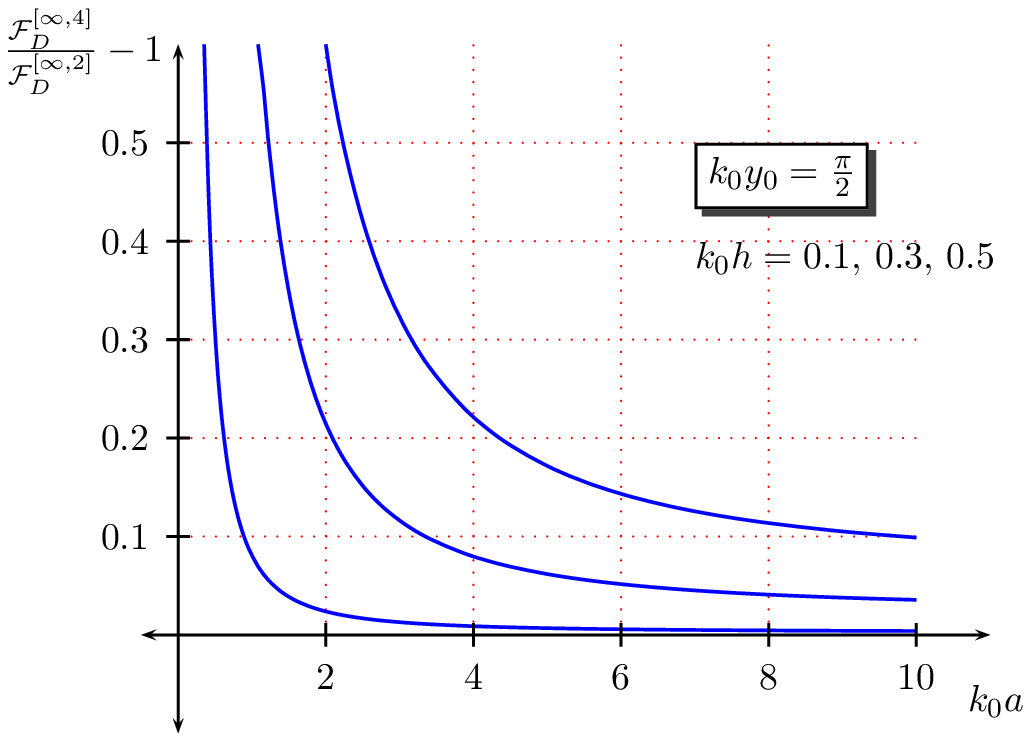}
\\
\includegraphics[width=80mm]{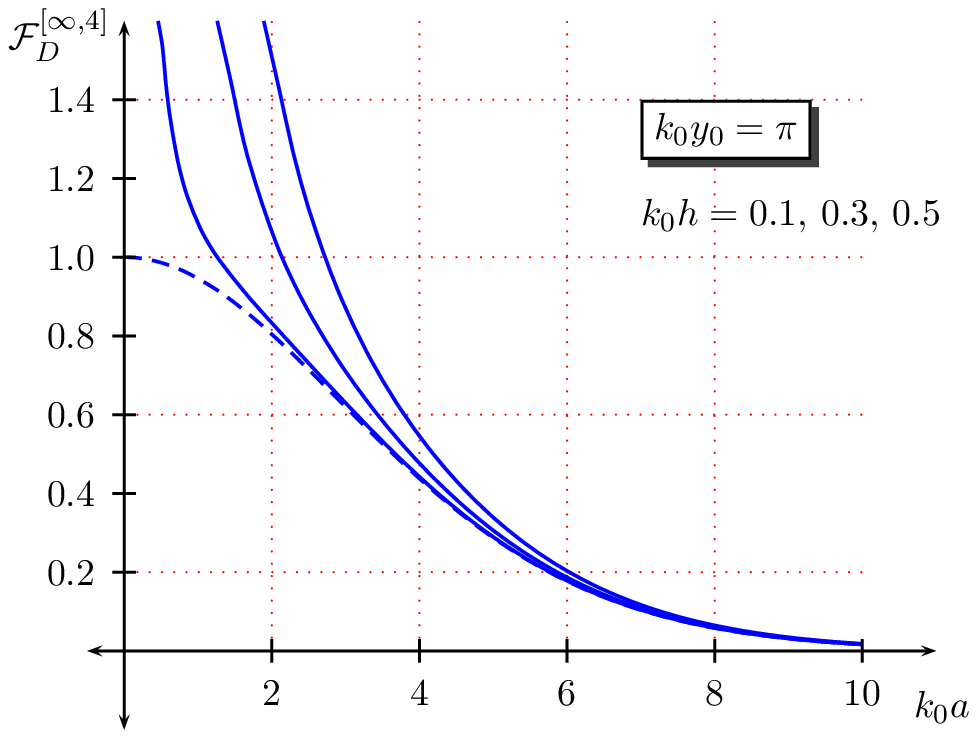} &
\includegraphics[width=80mm]{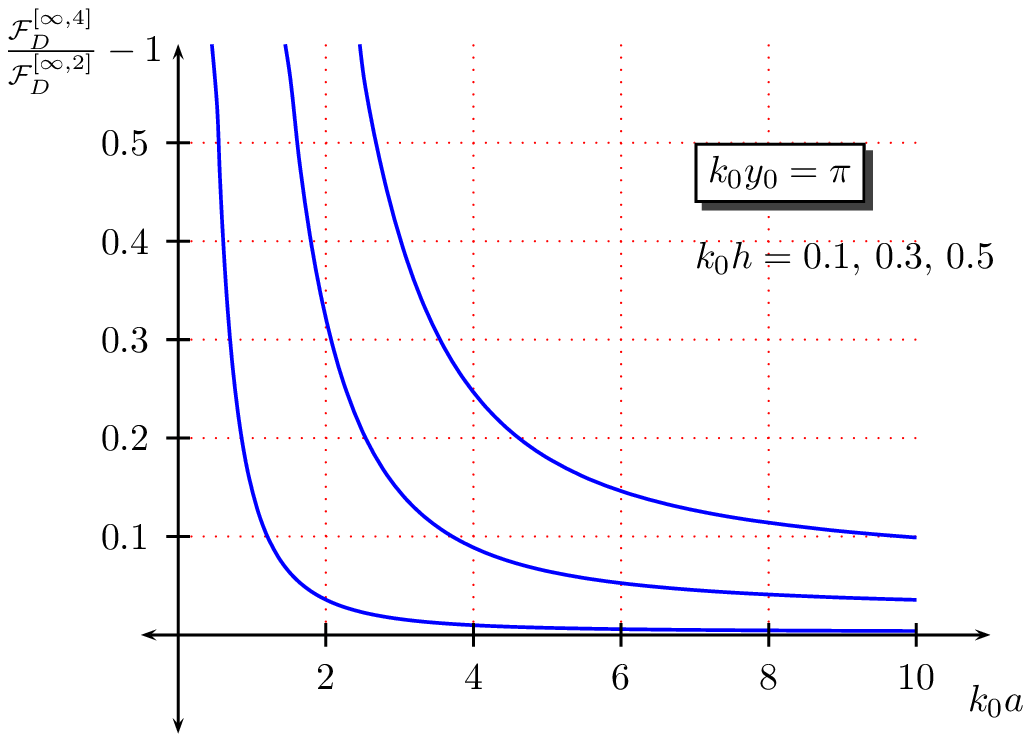}
\end{tabular}
\caption{ 
Dirichlet limit:
The plots on the left show ${\cal F}_D^{[\infty,4]}$ versus $k_0a$ for various
values of $k_0y_0$. The dashed curve represents ${\cal F}_D^{[\infty,2]}$
which is plotted here for reference.
In each plot the higher values of $k_0h$ deviate more from the reference.
The plots on the right show the fractional correction in the
next-to-leading order contribution.
In each plot the higher values of $k_0h$ have larger corrections.  }
\label{f4-D-fig}
\end{figure}

We plot the lateral force for the Dirichlet limit, 
${\cal F}_D^{[\infty,4]}$,
with the fractional corrections due to the next-to-leading contribution,
in figure \ref{f4-D-fig}. 
The plots diverge as $k_0a \rightarrow 2 k_0h$ because the perturbation
series breaks down beyond that point.
The plots for $k_0y_0=0$ do not diverge, see eq.~\eqref{latf-frac}.
We observe that, while keeping $k_0y_0$ fixed, the fractional 
corrections due to the next-to-leading order are small
for $k_0h \ll 1$.

In the absence of an exact answer, for the Dirichlet case, 
to compare with it is not possible 
to extract the precise error in the perturbative results. 
In the next section we shall evaluate the corresponding result in 
the weak coupling limit and evaluate the lateral Casimir force 
non-perturbatively. We show that the error in the lateral force,
for the weak coupling limit,
after the next-to-leading order has been included is 
small for $k_0h \ll 1$. Presuming that the same would
hold for the Dirichlet case, we can expect the error 
to be sufficiently small for comparison with experimental data. 
(Comparison with experiments requires the corresponding 
evaluation for the electromagnetic case, 
which will be taken up in a later publication.)

\subsubsection{Weak coupling limit - Non-perturbative results}

\begin{figure}
\begin{tabular}{cc}
\includegraphics[width=80mm]{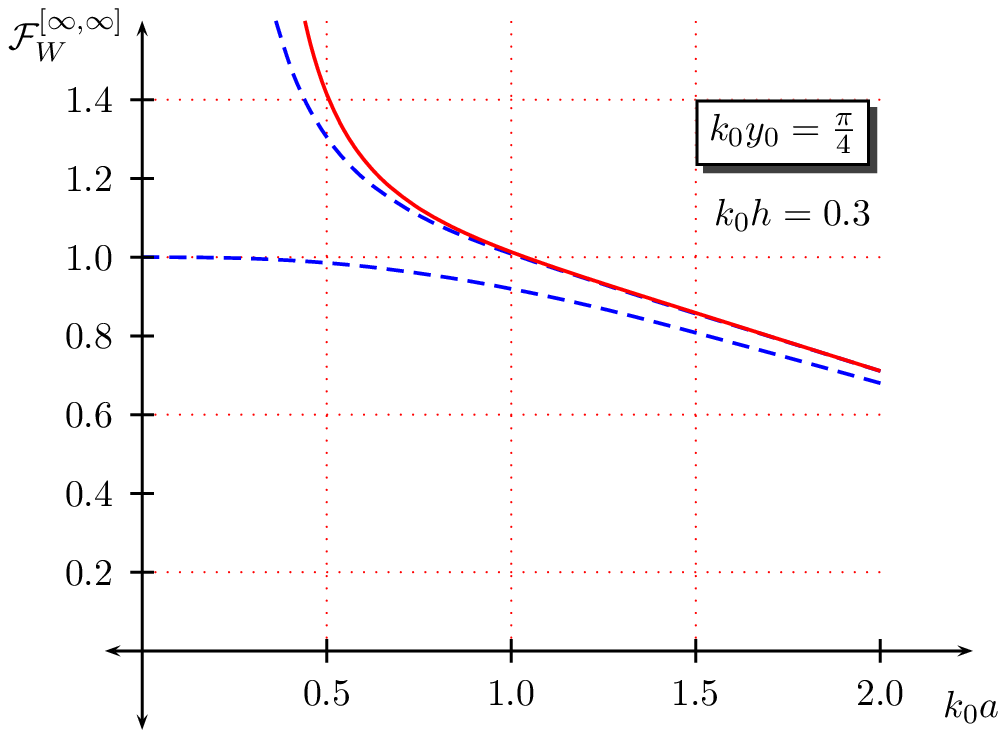} &
\includegraphics[width=80mm]{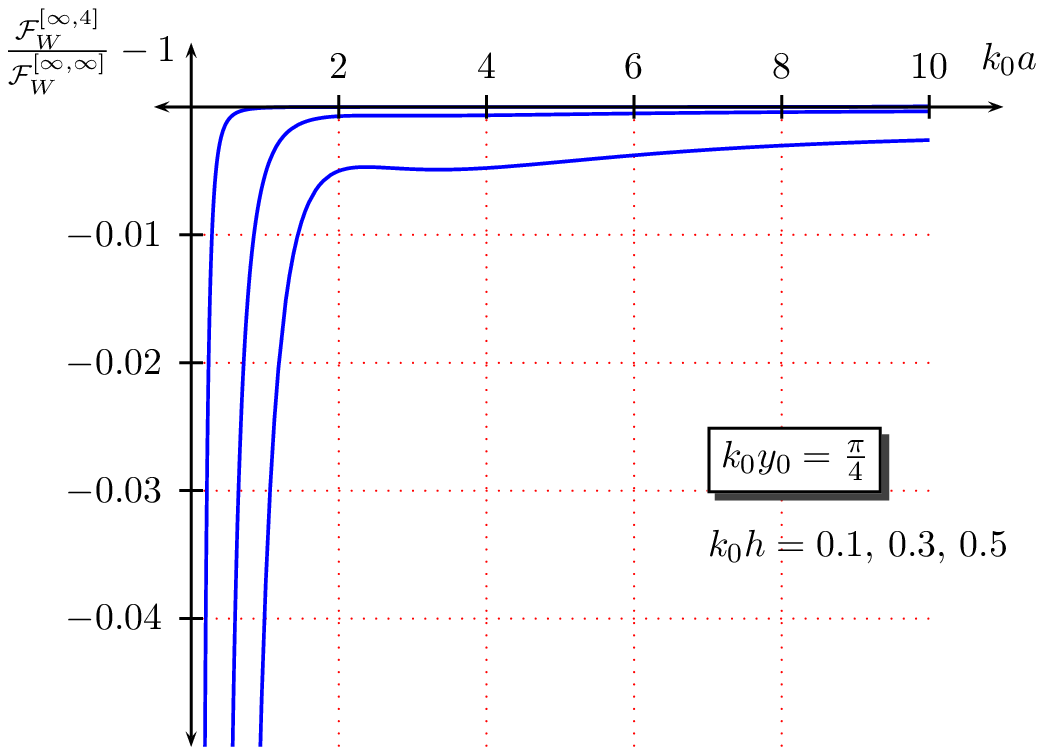}
\\
\includegraphics[width=80mm]{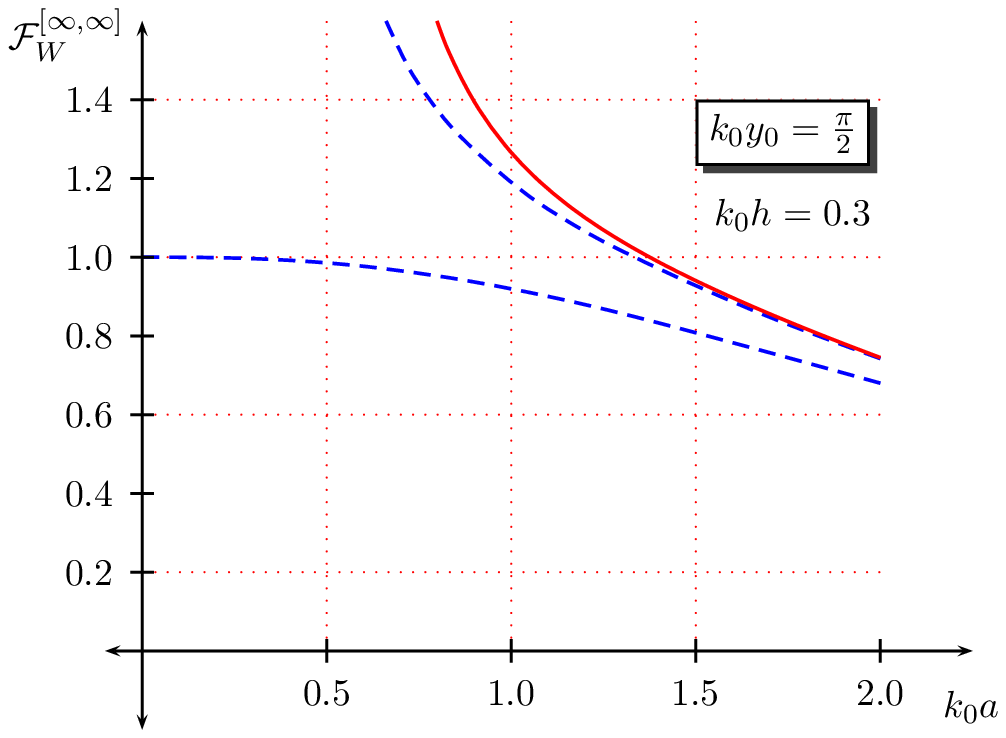} &
\includegraphics[width=80mm]{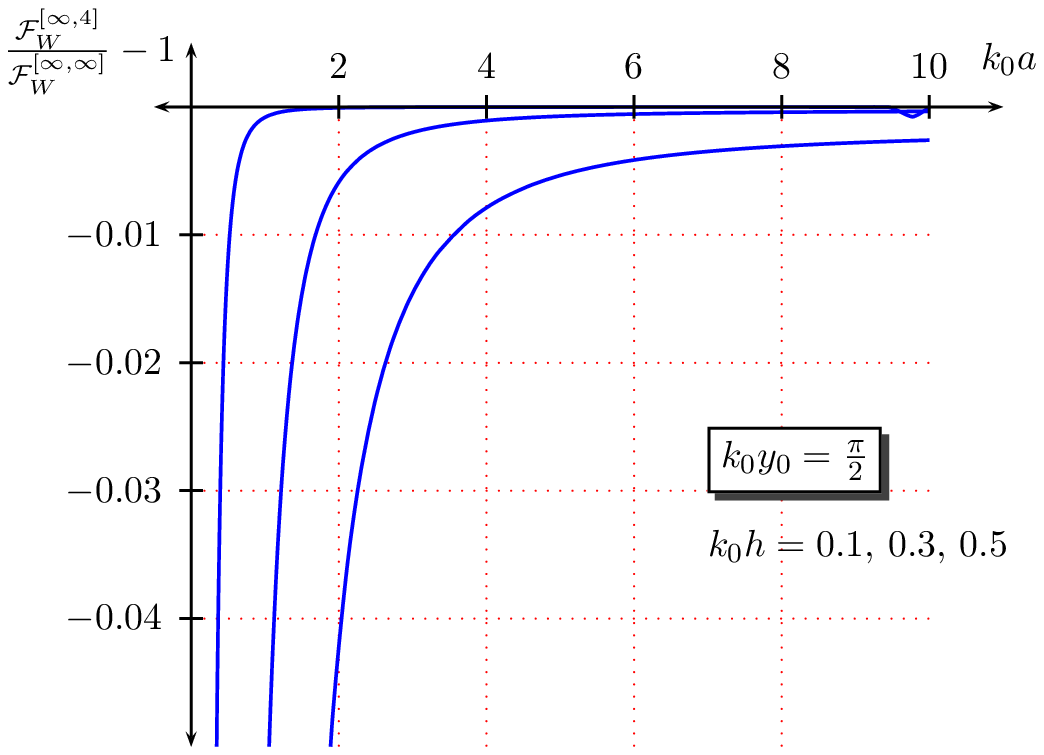}
\\
\includegraphics[width=80mm]{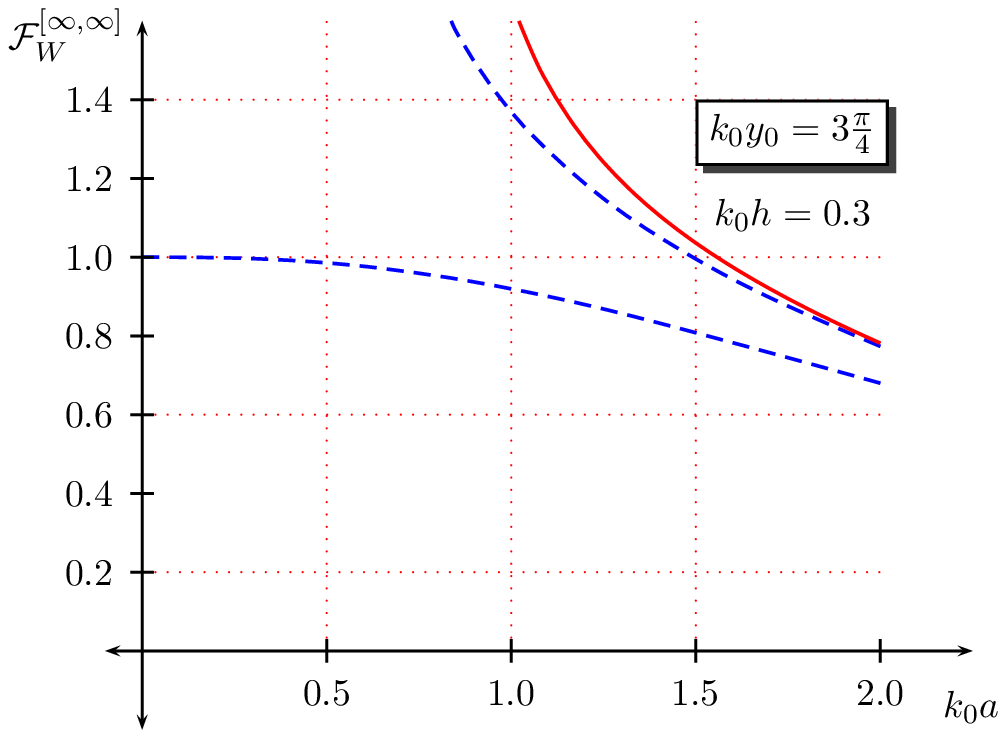} &
\includegraphics[width=80mm]{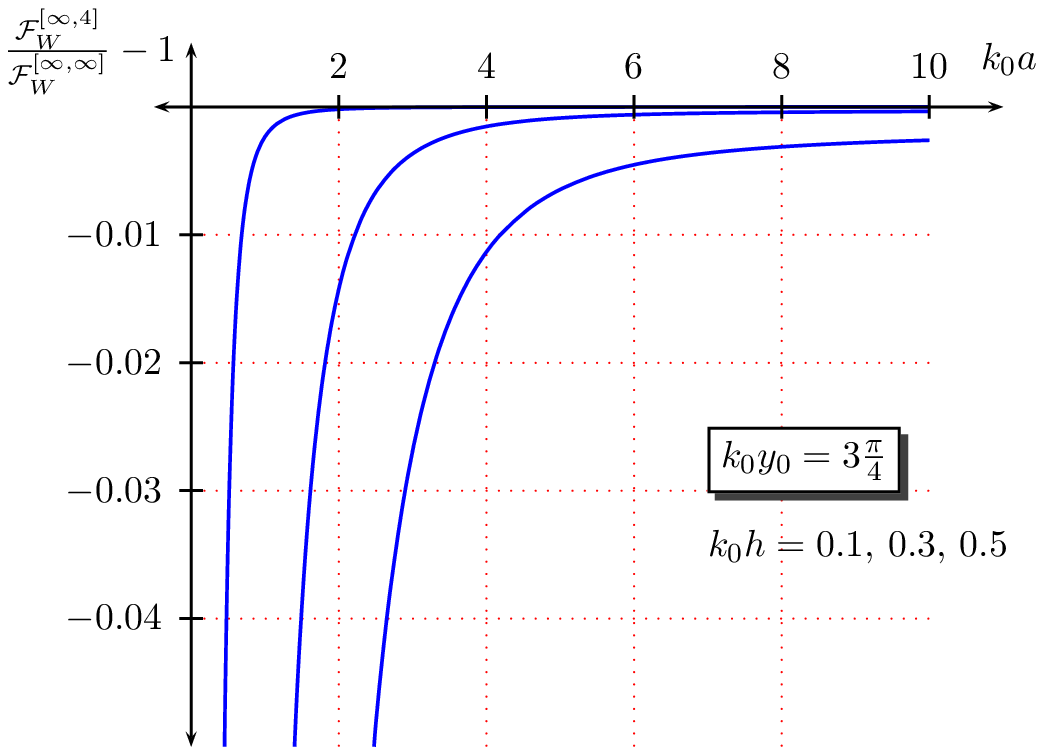}
\end{tabular}
\caption{
Weak coupling limit:
The plots on the left show ${\cal F}_D^{[\infty,\infty]}$ versus $k_0a$ 
for various values of $k_0y_0$ at $k_0h=0.3$.
The dashed curves represents
${\cal F}_D^{[\infty,2]}$ and ${\cal F}_D^{[\infty,4]}$
which is plotted here for reference.
In each plot the exact value is greater than the value estimated
by ${\cal F}_D^{[\infty,4]}$.
The plots on the right show the fractional error in the perturbative result.
In each plot the higher values of $k_0h$ have larger errors.  }
\label{exact-W-fig}
\end{figure}

The plots for the lateral force in the weak coupling limit,
${\cal F}_W^{[\infty,4]}$,
which have not been displayed here, have the same qualitative content
as the Dirichlet case shown in figure \ref{f4-D-fig}.
To estimate the error in the perturbative results we shall now 
attempt a non-perturbative evaluation of the lateral Casimir force 
in the weak coupling limit.

We begin by pointing out the remarkable observation
in~\cite{Milton:2007wz} that exact results for the Casimir force,
in the weak coupling limit, can be achieved for specific geometries.
This is being extended for a class of geometries,
for the scalar case, in~\cite{Milton:}, and another paper will 
cover dielectrics in electromagnetism, in \cite{Milton:2008vr}.
Starting from the central formula in~\cite{Milton:},
(alternatively using eq.~\eqref{formal-series}, 
for the weak coupling limit,) we will have
\begin{equation}
\frac{E_W}{L_x} = - \frac{\lambda_1 \lambda_2}{32\pi^3}
\int_{-\infty}^\infty dy \int_{-\infty}^\infty dy^\prime
\frac{1}{(y-y^\prime)^2 +a(y,y^\prime)^2},
\end{equation}
where 
$a(y,y^\prime) = a + h_2 \sin [k_0 y^\prime] - h_1 \sin [k_0 (y + y_0)]$.
Changing variables,
$y-y^\prime \rightarrow y$, $y + y^\prime \rightarrow 2 \theta/k_0$,
and making substitutions very similar to those made after 
eq.~\eqref{EDPFA}, we have
\begin{equation}
\frac{E_W}{L_xL_y} = - \frac{\lambda_1 \lambda_2}{32\pi^2}
\frac{1}{2\pi^2} \int_{-\infty}^\infty dy \int_0^{2\pi} d\theta
\frac{1}{y^2 +[a - r(k_0y) \cos\theta]^2},
\label{EW=intyth}
\end{equation}
where $r(k_0y)^2 = h_1^2 + h_2^2 - 2 h_1 h_2 \cos (k_0y + k_0y_0)$,
which is different from the expression for $r$ in the PFA case,
presented after eq.~\eqref{EDPFA}, in the dependence in the 
argument of the cosine function.
Nevertheless, the $\theta$ integral can still be performed using
eq.~\eqref{a+rcos-int} to yield
\begin{equation}
\frac{E_W}{L_xL_y} = - \frac{\lambda_1 \lambda_2}{32\pi^2}
\frac{1}{\pi} \int_{-\infty}^\infty \frac{dy}{y} 
\,\text{Re} \bigg[ \frac{1}{(y + ia)^2 + r(k_0y)^2} \bigg]^\frac{1}{2},
\quad \text{for} \quad |h_1| + |h_2| < a.
\end{equation}
Using the above expression in eq.~\eqref{latf} we calculate the lateral force
per unit area in the weak coupling limit to be
\begin{equation}
\frac{F_\text{Lat,W}}{L_xL_y} = - \frac{\lambda_1 \lambda_2}{32\pi^2\,a^2}
\frac{h_1}{a} \frac{h_2}{a} (k_0a)^4
\frac{1}{\pi} \int_{-\infty}^\infty \frac{dt}{t}
\,\text{Re} \bigg[ \frac{\sin (t + k_0y_0)}
            {[(t + ik_0a)^2 + \{k_0 r(t)\}^2]^\frac{3}{2}} \bigg],
\quad \text{for} \quad |h_1| + |h_2| < a.
\label{FW-Re-int}
\end{equation} 
Expanding the denominator in the above expression as a power series
in $k_0r/(t + i k_0a)$, lets us evaluate the integral using the 
residue theorem. We can thus write, after letting $h_1=h_2 =h$, 
which leads to the simplification, $r(k_0y_0) = 2 h \sin[(k_0y + k_0y_0)/2]$,
\begin{equation}
\frac{F_\text{Lat,W}}{L_xL_y} = \frac{\lambda_1 \lambda_2}{32\pi^2\,a^2}
(k_0a)^2 \text{Re} \bigg[ \frac{e^{-ik_0y_0}}{(-i)}
\sum_{n=0}^\infty \frac{1}{[n!]^2} \frac{1}{2n+2}
\Big( h \frac{\partial}{\partial a} \Big)^{2n+2}
\frac{1}{k_0a} \,e^{-k_0a} 
\sin^{2n} \Big( \frac{k_0y_0}{2} - i \frac{k_0a}{2} \Big) \bigg],
\label{FW-sum}
\end{equation}
which reproduces the leading and next-to-leading results in
eqs.~\eqref{latf-2W} and \eqref{latf-4W} for the case
$n=0$ and $n=1$ respectively.
This structure justifies our presumption that the perturbative
analysis holds as an asymptotic expansion as $k_0h \to 0$.
Expressing the above sum in a closed form, or evaluation of 
the integral in eq.~\eqref{FW-Re-int}, would lead to an exact 
expression for the lateral force in the weak coupling limit.
We have not been able to achieve either, however, so we shall 
rely on numerical evaluation.

In terms of the notation introduced in eq.~\eqref{dless-force}
we can write eq.~\eqref{FW-Re-int} as
\begin{equation}
{\cal F}_W^{[\infty,\infty]} =
- \frac{(k_0a)^3}{\sin (k_0y_0)} \frac{1}{\pi} 
\int_{-\infty}^\infty \frac{dt}{t}
\,\text{Re} \bigg[ \frac{\sin (t + k_0y_0)}
            {[(t + ik_0a)^2 + \{k_0 r(t)\}^2]^\frac{3}{2}} \bigg],
\quad \text{for} \quad 2h < a.
\label{FW=intt}
\end{equation}
The above non-perturbative expression for the 
lateral Casimir force on corrugated 
plates, in the weak coupling limit, has been plotted in 
figure~\ref{exact-W-fig}. For comparison purposes the perturbative 
results are plotted as dashed curves in figure~\ref{exact-W-fig}.
Actually, we found the numerical evaluation of eq.~\eqref{EW=intyth},
after evaluating the derivative with respect to $y_0$,
easier than that of eq.~\eqref{FW=intt}.
We observe that the perturbative results,
when the next-to-leading order is included, 
compares with the exact result remarkably well for 
$k_0h \ll 1$ and $2h < a$.
The fractional error in the lateral force when the next-to-leading
order contribution is included is displayed on the right column in
figure~\ref{exact-W-fig}.

\subsubsection{Validity of PFA}

\begin{figure}
\begin{tabular}{cc}
\includegraphics[width=75mm]{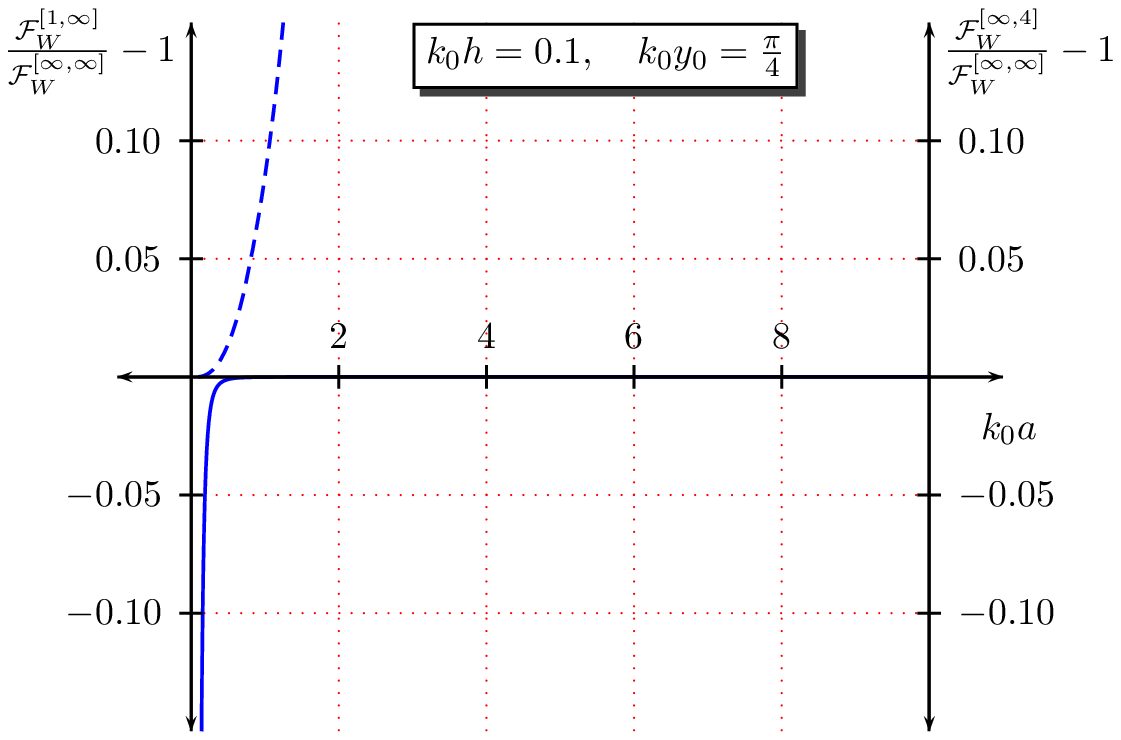} &
\hspace{5mm}
\includegraphics[width=75mm]{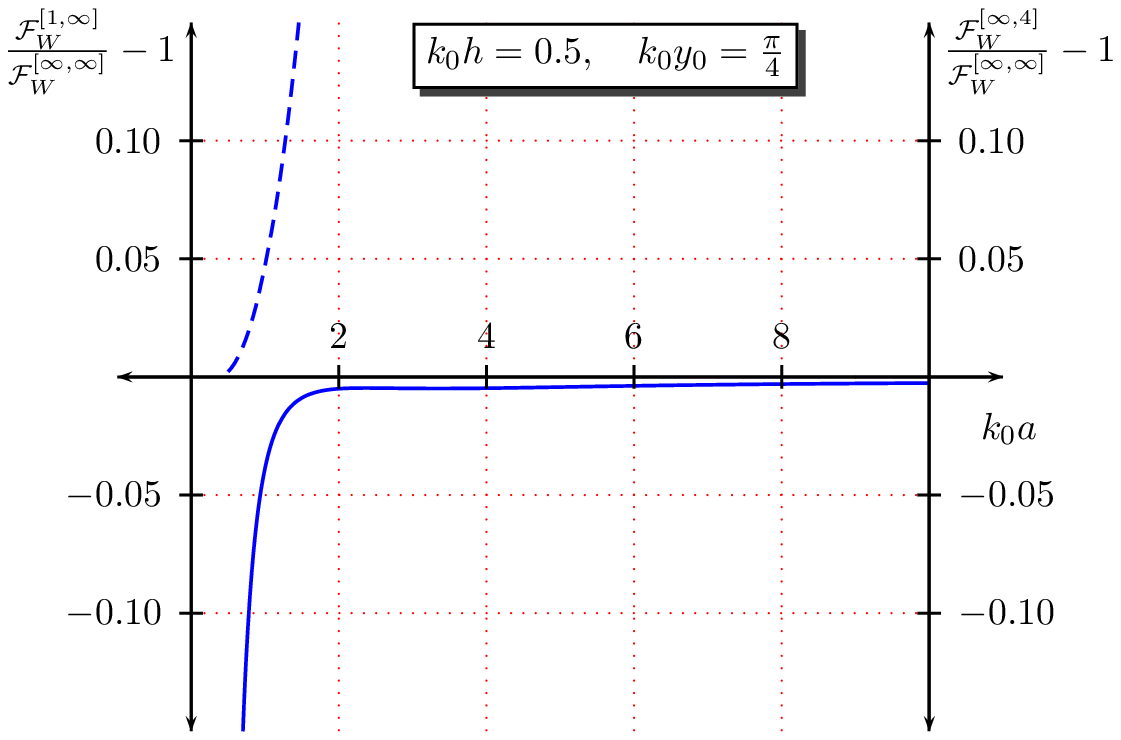} 
\end{tabular}
\caption{ Weak coupling limit:
Fractional error in PFA is plotted as the dashed curve.
The corresponding error in the perturbative result is plotted 
as the bold curve and is described by the axis on the right.}
\label{pfa-and-exact-W}
\end{figure}

We shall now compare the PFA result in the weak limit, 
derived in eq.~\eqref{latf-PFA-W}, with the non-perturbative result
in eq.~\eqref{FW-Re-int}.
In figure~\ref{pfa-and-exact-W} we plot the fractional error 
in the PFA result, for fixed $k_0h$, and compare it to the 
fractional error in the perturbative results.
We note that PFA is a good approximation for $k_0a \ll 1$ 
and perturbative results are valid for $k_0h \ll 1$.
Both the PFA and the perturbative analysis are restricted to $2h<a$.
The $2h<a$ restriction is necessary for the force being finite,
due to non-overlapping of the potentials for arbitrary offset.
For special offsets, the restriction is weaker, and there is no 
restriction at all, for zero offset. The precise dependence of
the restriction on the offset is gotten by referring to eq.~\eqref{latf-PFA-W}
and observing that the validity regime is $r = 2h\sin (k_0y_0/2) < a$. 

\begin{figure}
\begin{tabular}{cc}
\includegraphics[width=80mm]{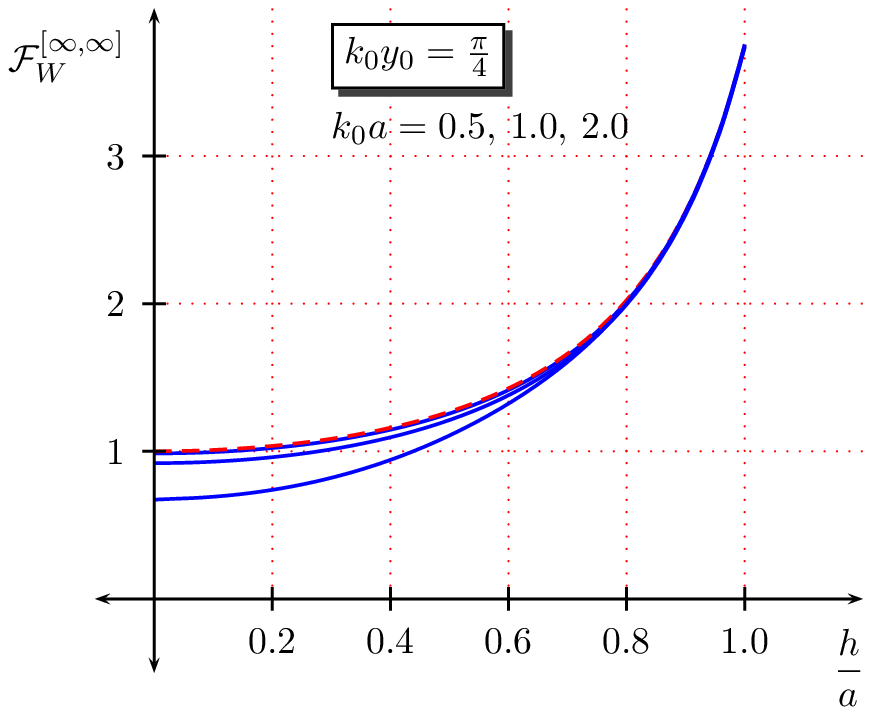} &
\hspace{5mm}
\includegraphics[width=80mm]{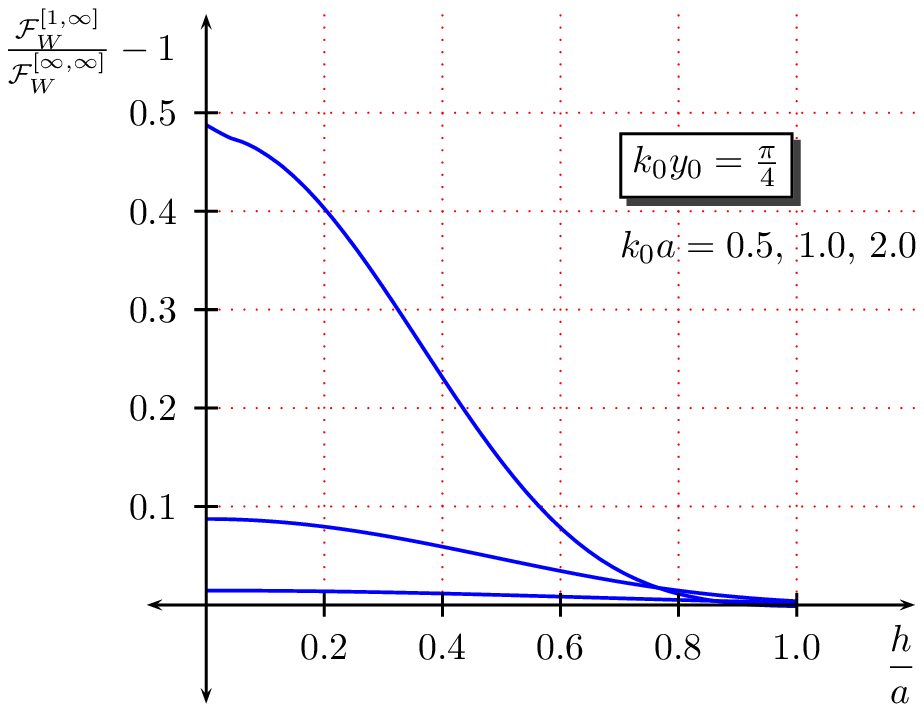}
\end{tabular}
\caption{ Weak coupling limit:
On the left the PFA is shown as the dashed curve. The exact results for 
higher values of $k_0a$ are seen to deviate from the dashed curve.
On the right the fractional error in PFA is shown to increase
with higher values of $k_0a$.}
\label{pfa-and-exact-W-versus-hoa}
\end{figure}
A recent debate, see \cite{Rodrigues:2006, Chen:2007, Rodrigues:2007},
involved the comparison of the PFA result and the perturbative result.
We earlier noted that the PFA limit is obtained by taking the
limit $k_0a \rightarrow 0$ while keeping the ratio 
$h/a$ fixed. In figure~\ref{pfa-and-exact-W-versus-hoa}
we plot the lateral force in the PFA limit and compare it 
with the non-perturbative result for various values of $k_0a$.
This justifies our presumption that the PFA is a good approximation
for $k_0a \ll 1$.
We note that the error in the PFA is less than 1\% for $k_0a \ll 1$
for arbitrary $h/a$.
For $k_0y_0 = \pi/4$, we observe that the PFA is a very good
approximation for $h\sim a$ which satisfies $2h\sin (k_0y_0/2) < a$.
After viewing the plots for various values of $k_0y_0$ we note that 
in general the PFA is a good approximation for $h\sim a$ and 
further beyond for offsets $k_0y_0 < \pi/4$.
It is, in fact, plausible that the PFA holds for large amplitude
corrugations for small offsets because the corrugations fit together
like fingers in a glove.

\begin{acknowledgments}
We thank Jef Wagner for extensive collaborative assistance
throughout this project.
KVS would like to thank Osama Alkhouli, Subrata Bal, Pravin Chaubey,
James Dizikes, David Hartnett, K. V. Jupesh, Sai Krishna Rao, S. Shankar,
and St\'{e}phane Valladier, for discussions, comments, 
and help with programming in Mathematica.
We thank Steve Fulling for constructive comments on the manuscript.
We thank the US National Science Foundation (Grant No. PHY-0554926)
and the US Department of Energy (Grant No. DE-FG02-04ER41305)
for partially funding this research.
ICP would like to thank the French National Research Agency (ANR) 
for support through Carnot funding.
\end{acknowledgments}
\appendix
\section{Derivatives of the Green's function}
\label{ddz-green}

Evaluation of eq.~\eqref{I2} involves taking derivatives of 
the Green's function which as we shall see involves evaluating a
function at a point where it has jump discontinuities
in the derivatives.
We start by noting that the differential equation in eq.~\eqref{g0zzp} 
can be solved in terms of exponential functions 
in the different regions described in figure \ref{regions}. Explicitly
the solutions are given by the following piecewise-defined functions, 
with subscripts denoting the regions they are defined in,
\begin{subequations}
\begin{eqnarray}
g_1^{(0)}(z,z^\prime;\kappa)
&=& \frac{1}{2\kappa} \,e^{-\kappa |z-z^\prime|}
- \frac{1}{\Delta} \frac{1}{2\kappa}
\,e^{\kappa (z+z^\prime)}
\left[ \frac{\lambda_1}{2\kappa} \,e^{-2\kappa a_1}
+ \frac{\lambda_2}{2\kappa} \,e^{-2\kappa a_2}
- \frac{\lambda_1}{2\kappa} \frac{\lambda_2}{2\kappa}
\Big( e^{-2\kappa a_2} - e^{-2\kappa a_1} \Big) \right],
\\
g_2^{(0)}(z,z^\prime;\kappa)
&=& \frac{1}{2\kappa} \,e^{-\kappa |z-z^\prime|}
- \frac{1}{\Delta} \frac{1}{2\kappa}
\left[ \frac{\lambda_2}{2\kappa} \Big( 1 + \frac{\lambda_1}{2\kappa} \Big)
\,e^{\kappa (z + z^\prime - 2a_2)}
+ \frac{\lambda_1}{2\kappa} \Big( 1 + \frac{\lambda_2}{2\kappa} \Big)
\,e^{-\kappa (z + z^\prime - 2a_1)} \right.
\nonumber \\
&& \hspace{35mm}
\left.
- \frac{\lambda_1}{2\kappa} \frac{\lambda_2}{2\kappa}
\,e^{\kappa (z - z^\prime - 2a)}
- \frac{\lambda_1}{2\kappa} \frac{\lambda_2}{2\kappa}
\,e^{-\kappa (z - z^\prime + 2a)} \right],
\label{g20zzp-sol}
\\
g_3^{(0)}(z,z^\prime;\kappa)
&=& \frac{1}{2\kappa} \,e^{-\kappa |z-z^\prime|}
- \frac{1}{\Delta} \frac{1}{2\kappa}
\,e^{-\kappa (z+z^\prime)}
\left[ \frac{\lambda_1}{2\kappa} \,e^{2\kappa a_1}
+ \frac{\lambda_2}{2\kappa} \,e^{2\kappa a_2}
- \frac{\lambda_1}{2\kappa} \frac{\lambda_2}{2\kappa}
\Big( e^{2\kappa a_1} - e^{2\kappa a_2} \Big) \right],
\\ 
g_4^{(0)}(z,z^\prime;\kappa)
&=& \frac{1}{\Delta} \frac{1}{2\kappa}
\left[ \Big( 1 + \frac{\lambda_2}{2\kappa} \Big) \,e^{\kappa (z - z^\prime)}
- \frac{\lambda_2}{2\kappa} \,e^{\kappa (z + z^\prime - 2a_2)} \right],
\\
g_5^{(0)}(z,z^\prime;\kappa)
&=& \frac{1}{\Delta} \frac{1}{2\kappa} \,e^{\kappa (z-z^\prime)},
\\
g_7^{(0)}(z,z^\prime;\kappa)
&=& \frac{1}{\Delta} \frac{1}{2\kappa}
\left[ \Big( 1 + \frac{\lambda_1}{2\kappa} \Big) \,e^{\kappa (z - z^\prime)}
- \frac{\lambda_1}{2\kappa} \,e^{-\kappa (z + z^\prime - 2a_1)} \right],
\end{eqnarray}\label{g0zzp-sol}\end{subequations}
where
\begin{equation}
\Delta = 1 + \frac{\lambda_1}{2\kappa} + \frac{\lambda_2}{2\kappa}
+ \frac{\lambda_1}{2\kappa} \frac{\lambda_2}{2\kappa}
\,\Big(1 - e^{-2\kappa a} \Big).
\label{Delta}
\end{equation}
Using the reciprocal symmetry of the Green's function we further have 
$ g_6^{(0)}(z,z^\prime;\kappa) = g_4^{(0)}(z^\prime,z;\kappa)$,
$ g_8^{(0)}(z,z^\prime;\kappa) = g_5^{(0)}(z^\prime,z;\kappa)$,
and $ g_9^{(0)}(z,z^\prime;\kappa) = g_7^{(0)}(z^\prime,z;\kappa)$.
The above piecewise solution gives the complete Green's function.

\begin{figure}
\begin{center}
\includegraphics[width=50mm]{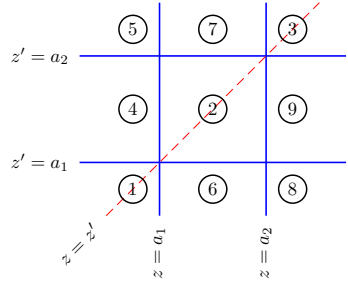}
\caption{Description of the regions in the $z-z^\prime$ domain 
on which the piecewise functions contributing to the Green's function
are defined.}
\label{regions}
\end{center}
\end{figure}

It is of relevance to observe that the Green's function represented 
above is continuous everywhere in the $z,z^\prime$ domain while
its first derivatives have simple (jump) discontinuities along the lines, 
$z=z^\prime$, $z=a_{1,2}$, and $z^\prime = a_{1,2}$.
Since the evaluation of $I$-kernels
in eq.~\eqref{I2}, and in eq.~\eqref{Imi}, 
involves taking derivatives of the Green's 
function at the lines of discontinuity, a definite 
prescription for the evaluation of the derivatives is called for.
For this purpose, we interpret the value of a function at the point
where it has a jump discontinuity to be the average of all the 
possible limiting values. For example, the derivative of the 
Green's function at the point $(z=a_2,z^\prime=a_1)$
in figure \ref{regions} takes on four different values when approached from 
the four regions 2, 6, 8, and 9. The value of the derivative is 
defined as the average of these four values. 

We provide the following supporting arguments for the averaging 
prescription.
Firstly, we note that eq.~\eqref{I2} was derived from eq~\eqref{dE12-2}
after integrating over the complete $z-z^\prime$ domain. 
The potentials in eq~\eqref{dE12-2} involved derivatives
of delta functions, and after integration by parts the integrals
got contributions from a single point in the $z-z^\prime$ domain.
Since an integral is a sum we expect it to support the averaging
prescription. The second justification we provide is in the way 
of verification for known examples. 
We made checks by using this prescription to analyze the Green's
function for the case of a single plate. We do not provide the 
details of the exercise here. 

Evaluation of the derivatives using the averaging prescription
is made convenient when we define the Green's function around a
point as the average of the suitable piecewise functions.
Thus, we evaluate
\begin{subequations}
\begin{eqnarray}
g^{(0)}(a_1,a_2;\kappa)
&=& \frac{1}{4} \Big[ g_2^{(0)}(a_1,a_2;\kappa) + g_4^{(0)}(a_1,a_2;\kappa)
+ g_5^{(0)}(a_1,a_2;\kappa) + g_7^{(0)}(a_1,a_2;\kappa) \Big]
= \frac{1}{\Delta} \frac{1}{2\kappa} \,e^{-\kappa a},
\nonumber \\
g^{(0)}(a_2,a_1;\kappa) 
&=& \frac{1}{4} \Big[ g_2^{(0)}(a_2,a_1;\kappa) + g_6^{(0)}(a_2,a_1;\kappa)
+ g_8^{(0)}(a_2,a_1;\kappa) + g_9^{(0)}(a_2,a_1;\kappa) \Big]
= \frac{1}{\Delta} \frac{1}{2\kappa} \,e^{-\kappa a},
\nonumber \\
g^{(0)}(a_1,a_1;\kappa)
&=& \frac{1}{4} \Big[ g_1^{(0)}(a_1,a_1;\kappa) + g_2^{(0)}(a_1,a_1;\kappa)
+ g_4^{(0)}(a_1,a_1;\kappa) + g_6^{(0)}(a_1,a_1;\kappa) \Big]
= \frac{1}{\Delta} \frac{1}{2\kappa}
\left[ 1 + \frac{\lambda_2}{2\kappa} 
\,\Big( 1 - e^{-2\kappa a} \Big)  \right],
\nonumber \\
g^{(0)}(a_2,a_2;\kappa)
&=& \frac{1}{4} \Big[ g_2^{(0)}(a_2,a_2;\kappa) + g_3^{(0)}(a_2,a_2;\kappa)
+ g_7^{(0)}(a_2,a_2;\kappa) + g_9^{(0)}(a_2,a_2;\kappa) \Big]
= \frac{1}{\Delta} \frac{1}{2\kappa}
\left[ 1 + \frac{\lambda_1}{2\kappa}
\,\Big( 1 - e^{-2\kappa a} \Big)  \right].
\nonumber
\end{eqnarray}\end{subequations}
The first derivatives evaluate to be
\begin{subequations}
\begin{align}
\partial_z \,g^{(0)}(z,z^\prime;\kappa) \big|_{z=a_1,z^\prime=a_2}
&= \frac{1}{\Delta} \frac{\kappa}{2\kappa} 
\Big( 1 + \frac{\lambda_1}{2\kappa} \Big) \,e^{-\kappa a},
\\
\partial_{z^\prime} \,g^{(0)}(z,z^\prime;\kappa) 
\big|_{z=a_1,z^\prime=a_2}
&= - \frac{1}{\Delta} \frac{\kappa}{2\kappa}
\Big( 1 + \frac{\lambda_2}{2\kappa} \Big) \,e^{-\kappa a},
\\
\partial_z \,g^{(0)}(z,z^\prime;\kappa) \big|_{z=a_1,z^\prime=a_1}
&= - \frac{1}{\Delta} \frac{\kappa}{2\kappa} 
\frac{\lambda_2}{2\kappa} \,e^{-\kappa a},
\\
\partial_z \,g^{(0)}(z,z^\prime;\kappa) \big|_{z=a_2,z^\prime=a_2}
&= \frac{1}{\Delta} \frac{\kappa}{2\kappa}
\frac{\lambda_1}{2\kappa} \,e^{-\kappa a}.
\end{align}\end{subequations}
where $\partial_{z,z^\prime}$ is derivative with respect to $z,z^\prime$.
The second derivatives involving two distinct variables evaluate to be
\begin{subequations}
\begin{eqnarray}
\partial_z \partial_{z^\prime} 
       \,g^{(0)}(z,z^\prime;\kappa) \big|_{z=a_1,z^\prime=a_2}
&=& - \frac{1}{\Delta} \frac{\kappa^2}{2\kappa}
\Big( 1 + \frac{\lambda_1}{2\kappa} \Big) 
\Big( 1 + \frac{\lambda_2}{2\kappa} \Big) \,e^{-\kappa a},
\\[2mm]
\partial_z \partial_{z^\prime}        
       \,g^{(0)}(z,z^\prime;\kappa) \big|_{z=a_1,z^\prime=a_1}
&=&
- \frac{1}{\Delta} \frac{\kappa^2}{2\kappa}
\left[ \Big( 1 + \frac{\lambda_1}{2\kappa} \Big)
\Big( 1 + \frac{\lambda_2}{2\kappa} \Big)
+ \frac{\lambda_2}{2\kappa} \,e^{-2 \kappa a} \right],
\\[2mm]
\partial_z \partial_{z^\prime}        
       \,g^{(0)}(z,z^\prime;\kappa) \big|_{z=a_2,z^\prime=a_2}
&=&
- \frac{1}{\Delta} \frac{\kappa^2}{2\kappa}
\left[ \Big( 1 + \frac{\lambda_1}{2\kappa} \Big)
\Big( 1 + \frac{\lambda_2}{2\kappa} \Big)
+ \frac{\lambda_1}{2\kappa} \,e^{-2 \kappa a} \right].
\end{eqnarray}\end{subequations}
The second derivatives involving the same variables evaluate to
\begin{equation}
\partial_z^2 g^{(0)}(z,z^\prime;\kappa) 
= \kappa^2 \,g^{(0)}(z,z^\prime;\kappa)
\end{equation}
which when compared with eq.~\eqref{g0zzp} tells us that the 
averaging prescription practically throws away the delta function
contributions in eq.~\eqref{g0zzp}. 
A more detailed study of this issue is being sought.
Since the second derivative returns the Green's function back,
all higher order derivatives are obtained in terms of the 
above expressions.

\section{Comparison of our leading order result with 
         reference \cite{Emig:2003}}
\label{app-emig}

Here we wish to explicitly compare
our results\textemdash 
in the leading order\textemdash with those in \cite{Emig:2003}.
In particular we show how the Dirichlet limit
of our expression for the interaction energy in the leading order
[eq.~\eqref{DE12=L2} with eq.~\eqref{L2D} inserted]
matches with the results in \cite{Emig:2003}
[eq.~(11) with eqs.~(13) and (17) there]. 
To this end we begin from eq.~\eqref{dE12-2}
and write the interaction energy in the form
\begin{equation}
E_{12}^{(2)}
= - \int d(t-t^\prime) \int\int dx\,dy \int\int dx^\prime dy^\prime
\,h_1(y) Q(|{\bf y}-{\bf y}^\prime|) h_2(y^\prime),
\label{E12-emig}
\end{equation}
where we have used the notation introduced in \cite{Emig:2003}:
${\bf y} \equiv (t,x,y)$,
$|{\bf y}-{\bf y}^\prime|^2 
= (t-t^\prime)^2 + (x-x^\prime)^2 + (y-y^\prime)^2$,
and switched to Euclidean time by replacing
$t,t^\prime,\tau \rightarrow -it,-it^\prime,-i\tau$.
The $Q$-kernel introduced above, following \cite{Emig:2003}, is given as 
\begin{equation}
Q(|{\bf y}-{\bf y}^\prime|)
= \frac{\lambda_1\lambda_2}{2}
\frac{\partial}{\partial z} \frac{\partial}{\partial z^\prime}
\Big[ G^{(0)}_E({\bf x},t,{\bf x}^\prime,t^\prime)
\,G^{(0)}_E({\bf x}^\prime,t^\prime,{\bf x},t) \Big] 
\bigg|_{z^\prime=a_1,z=a_2},
\label{Q=GG}
\end{equation}
where ${\bf x} \equiv (x,y,z)$ and 
$G^{(0)}_E({\bf x},t,{\bf x}^\prime,t^\prime)$
is given in terms of eq.~\eqref{G0zzp} after switching to Euclidean time as
\begin{equation}
G^{(0)}_E({\bf x},t,{\bf x}^\prime,t^\prime)
= \frac{1}{i} \, G^{(0)}({\bf x},t,{\bf x}^\prime,t^\prime)
= \int \frac{d^3\kappa}{(2\pi)^3} 
\,e^{i {\bm \kappa}\cdot ({\bf y} -{\bf y}^\prime)}
\,g^{(0)}(z,z^\prime;\kappa),
\label{emig-G}
\end{equation}
where ${\bm \kappa} \equiv (\zeta,k_x,k_y)$
and ${\bf y}=(t,x,y)$, ${\bf y}^\prime=(t^\prime,x^\prime,y^\prime)$.

\subsubsection*{Dirichlet Limit}

In the Dirichlet limit the above expression should correspond 
to the $Q$-kernel introduced in \cite{Emig:2003} [eq.~(13) there].
Note that in \cite{Emig:2003}
the interaction energy is not isolated from the outset unlike in our 
eq.~\eqref{DE12}. This is achieved in \cite{Emig:2003} by 
identifying and subtracting the contributions to the energy 
from the single plates.
We identify the presence of our $I$-kernel [see eq.~\eqref{I2}]
in eq.~\eqref{Q=GG} which lets us express the Dirichlet limit 
of the $Q$-kernel, using eq.~\eqref{I2D}, in the form
\begin{equation}
Q_D(|{\bf y}-{\bf y}^\prime|)
= \frac{1}{2} \bigg[ \int \frac{d^3\kappa}{(2\pi)^3}
\,e^{i {\bm \kappa}\cdot ({\bf y} -{\bf y}^\prime)}
\, \frac{\kappa}{\sinh \kappa a} \bigg]^2
= \frac{1}{2} \bigg[ \frac{1}{a^4} 
P\Big( \frac{|{\bf y}-{\bf y}^\prime|}{a} \Big) \bigg]^2
\end{equation}
in terms of the integral
\begin{equation}
P(x) = \frac{1}{2\pi^2} \frac{1}{x}
\int_0^\infty t^2\,dt \,\frac{\sin (t x)}{\sinh t}
= \frac{\pi}{8} \frac{1}{x} \frac{\sinh (\pi x/2)}{\cosh^3 (\pi x/2)}.
\end{equation}
This exact form for the $Q$-kernel in the Dirichlet limit
reproduces the result in \cite{Emig:2003}.

This exact form for the $Q$-kernel in the Dirichlet limit suggests 
that we can consider a class of corrugations for which exact results,
in the perturbative approximation, might be achievable. Like the exact
results achieved in \cite{Milton:,Milton:2008vr}
for the weak coupling limit,
we should be able to explore geometries in the Dirichlet limit in 
the perturbative approximation starting from eq.~\eqref{E12-emig}. 
Further, it should also be possible to extend
these explorations to the next-to-leading orders. 
And, for very special geometries it might just be possible 
to transcend the perturbative approximation and get exact results in the 
Dirichlet limit alone. We hope to be able to address some of these
explorations in our forthcoming publications.



\end{document}